\documentclass[11pt]{article}

\usepackage{graphicx}
\usepackage[footnotesize]{caption}

\usepackage{multirow}
\usepackage{dsfont}
\usepackage{amssymb}
\usepackage{slashed}
\usepackage{mathtools}
\usepackage{nicefrac}
\usepackage{verbatim}

\usepackage{axodraw}
\usepackage{color}

\newcommand{\rr}[1]{\mathrm{#1}}

\newcommand{\dd}[1]{\mathds{#1}}

\def\beq{\begin{equation}}
\def\eeq{\end{equation}}
\newcommand{\be}{\begin{eqnarray}}
\newcommand{\nn}{\nonumber}
\newcommand{\ee}{\end{eqnarray}}
\newcommand{\ba}{\begin{array}}
\newcommand{\ea}{\end{array}}
\newcommand{\ds}{\displaystyle}

\begin{document}
\begin{titlepage}

\vspace*{0.7cm}

\begin{center}
{
\bf\LARGE 750 GeV
Diphoton Resonance from Singlets in
an Exceptional Supersymmetric Standard Model
}
\\[12mm]
Stephen~F.~King$^{1,}$
\footnote{E-mail: \texttt{king@soton.ac.uk}},
Roman Nevzorov $^{2,}$\footnote{roman.nevzorov@adelaide.edu.au}
\\[-2mm]

\end{center}
\vspace*{0.50cm}
\centerline{$^{1}$ \it
Physics and Astronomy, University of Southampton,}
\centerline{\it
SO17 1BJ Southampton, United Kingdom }
\vspace*{0.2cm}
\centerline{$^{2}$
\it
ARC Centre of Excellence for Particle Physics at the Terascale,}
                  \centerline{\it
                  Department of Physics, The University of Adelaide,}
\centerline{\it
Adelaide, South Australia 5005, Australia}
\vspace*{1.20cm}

\begin{abstract}
\noindent
The 750-760 GeV diphoton resonance may be identified as one or two scalars and/or
one or two pseudoscalars
contained in the two singlet superfields $S_{1,2}$
arising from the three 27-dimensional representations of $E_6$.
The three 27s
also contain three copies of colour-triplet charge $\mp 1/3$ vector-like
fermions $D,\bar{D}$ and two copies of
charged inert Higgsinos $\tilde{H}^{+},\tilde{H}^{-}$
to which the singlets $S_{1,2}$ may couple.
We propose a variant of the E$_6$SSM
where the third singlet $S_3$ breaks a gauged $U(1)_N$ above the TeV scale,
predicting $Z'_N$, $D,\bar{D}$, $\tilde{H}^{+},\tilde{H}^{-}$
at LHC Run 2,
leaving the two lighter singlets $S_{1,2}$ with masses around 750 GeV.
We calculate the branching ratios and cross-sections for the two scalar and two pseudoscalar
states associated with the $S_{1,2}$
singlets, including possible degeneracies and maximal mixing,
subject to the constraint that their couplings remain perturbative
up to the unification scale.
\end{abstract}

\end{titlepage}

\thispagestyle{empty}
\vfill
\newpage

\section{Introduction}

Recently ATLAS and CMS experiments have reported  an excess of diphoton events at an invariant mass
around 750 GeV and 760 GeV from LHC Run-2 with $pp$ collisions at the center of mass energy of 13 TeV \cite{ATLAS,CMS}.
The  local significance of the excess ATLAS events is 3.9 $\sigma$ while that of the
excess CMS events is  2.6 $\sigma$, corresponding to the respective cross sections $\sigma(pp\to\gamma\gamma)=10.6$ fb
 and $\sigma(pp\to \gamma\gamma)=6.3$ fb.
ATLAS favours a broad width of $\Gamma \sim 45$ GeV, while CMS, although not excluding
a broad resonance, actually prefers a narrow width.
The diphoton excesses observed by ATLAS and CMS at this mass scale
may be partially understood by the factor of 5 gain in cross-section due to gluon production. However there is no evidence for any coupling of the resonance into anything except gluons and photons (no final states such as $f\bar{f}$, $VV$ ($f$ being a fermion and $V$ being
$W,Z$) since no missing $E_T$ or jets have been observed.

This may be the first indication of new physics at the TeV scale.
It could even be the tip of an iceberg of many future discoveries.
Several
interpretations have  been suggested based on extensions of the Standard Model spectrum ~\cite{Harigaya:2015ezk}-\cite{Karozas:2016hcp}.
Many of these papers suggest a spinless singlet coupled to vector-like fermions
\cite{Harigaya:2015ezk,Pilaftsis:2015ycr,Franceschini:2015kwy,Ellis:2015oso,Gupta:2015zzs,
Cao:2015pto,Kobakhidze:2015ldh,Falkowski:2015swt,Kim:2015ron,
Ding:2015rxx,Heckman:2015kqk,Kim:2015ksf,Cvetic:2015vit,Allanach:2015ixl,Wang:2015omi,Cai:2015hzc,Anchordoqui:2015jxc,Ibanez:2015uok,Karozas:2016hcp,Chao:2016mtn,Dutta:2016jqn,Palti:2016kew}.
Indeed, the observed resonance could be interpreted as a Standard Model scalar or pseudoscalar
singlet state $X$  with mass $m_X\sim 750-760$ GeV.
Moreover, because it decays into two photons,
its spin is consistent with $s=0$.
The process of generating the two photons can take place by the gluon-gluon fusion mechanism
according to the process $gg \to X\to \gamma\gamma $
hence it requires production and decay of the particle $X$. In a renormalisable theory
this interaction  can be  realised assuming vector-like fermions at the TeV scale,
which carry electric charge and colour.
Such vector-like pairs have not been observed at LHC, hence the mass of the fermion pair
should be around or above the TeV scale.
For example, in F-theory models based on $E_6$, low energy singlets coupling to
extra vector-like matter is predicted and
may be responsible for
the 750 GeV diphoton resonance \cite{Karozas:2016hcp}.
Such models motivate the phenomenological study of $E_6$ as being the origin of the new physics.

An example of a model with singlets and extra vector-like matter is the Exceptional Supersymmetric (SUSY) Standard Model
(E$_6$SSM)~\cite{King:2005jy,King:2005my},
where
the spectrum of the MSSM is extended to fill out three complete 27-dimensional representations
of the gauge group E$_6$ which is broken at the unification scale down to the SM gauge group
plus an additional gauged $U(1)_N$ symmetry at low energies under which right-handed neutrinos are neutral,
allowing them to get large masses.
The three $27_i$-plet families
(labelled by $i=1,2,3$) contain the usual quarks and leptons plus the following extra states:
SM-singlet fields, $S_i$; up- and down-type Higgs doublets, $H_{ui}$ and $H_{di}$;
and charged $\pm 1/3$ coloured, exotics $D_i$ and $\bar{D}_i$.
The extra matter ensures anomaly cancellation, however
the model also contains two extra SU(2) doublets, $H'$ and $\bar{H}'$, which are
required for gauge coupling unification~\cite{King:2007uj}.
To evade rapid proton decay a $\dd{Z}_2$ symmetry, either $\dd{Z}_2^{qq}$ or $\dd{Z}_2^{lq}$,
is introduced and to evade large flavour changing neutral currents an
approximate $\dd{Z}_2^H$ symmetry is introduced where
only the third family of Higgs doublets $H_{u3}$ and $H_{d3}$ and singlets $S_3$
are even under it and hence couple to
fermions and get vacuum expectation values (VEVs).
In particular, the third family singlet $S_3$ gets a VEV, $\langle S_3 \rangle = s/\sqrt{2}$,
which is responsible for the effective $\mu$ term,
inert Higgsino and D-fermion and $Z'_N$ masses, while
the first and second families of Higgs doublets and SM-singlets
do not get VEVs and are called ``inert''.
Further aspects of the theory and phenomenology
of this SUSY extension of the SM have been extensively studied in
\cite{Athron:2009bs,Athron:2009ue,Athron:2011wu,Accomando:2006ga,Hall:2010ix,
Hall:2011zq,Howl:2007zi,Callaghan:2011jj,Belyaev:2012jz}.

In this paper we take all three singlets to be even under the approximate $\dd{Z}_2^H$,
which allows them all to couple to
$\hat{H}_{ui}$ and $\hat{H}_{di}$ as well as $\hat{D}_i$ and $\hat{\bar{D}}_i$.
We shall assume that
the third singlet $S_3$ has appreciable couplings to the three families of
$H_{ui},H_{di}$ and $D_i$, $\bar{D}_i$, so that its large VEV generates effective mass
terms for all these states, as well a $Z'_N$, above the TeV scale
but possibly within the reach of LHC Run 2.
However
the first and second singlets $S_{1,2}$ may
have relatively small couplings to the third pair of Higgs doublets $H_{u3}$ and $H_{d3}$,
which are the only Higgs doublets to acquire VEVs.
In addition, we shall suppose that the value of the third singlet $S_3$ VEV $s$ is above the TeV scale,
while the other singlets $S_{1,2}$ at most develop small VEVs.
This is different from the modified E$_6$SSM in \cite{Chao:2016mtn},
where two of the singlets were assigned even under the approximate $\dd{Z}_2^H$
and both were allowed to develop VEVs and couple
to all three families of
$H_{ui},H_{di}$ and $D_i$, $\bar{D}_i$.
In the version of the E$_6$SSM here, we suppose that,
after the third singlet $S_3$ with large large $s$  VEV is integrated out, only
the first and second singlets $S_{1,2}$
appear in the low energy effective theory
and provide candidates for the 750-760 GeV resonance which may be
identfied as one or two scalars and/or
one or two pseudoscalars
contained in $S_{1,2}$.
The assumed smallness of the coupling of $S_{1,2}$ to $H_{u3}$ and $H_{d3}$
means that the observed resonance will not easily decay into pairs of top quarks or W bosons.

Many of the features of the considered model would be common to other SUSY E$_6$
models where the low energy spectrum consists of complete 27-plets. The present model is 
a variant of the E$_6$SSM and like that model is distinguished by the choice of surviving gauged 
$U(1)_N$ under which right-handed neutrinos have zero charge and may acquire large
Majorana masses, corresponding to a high scale seesaw mechanism. For earlier literature on 
other SUSY E$_6$ models based on different surviving gauged $U(1)$ symmetries under 
which right-handed neutrinos are charged see \cite{King:2005jy}.

The layout of the remainder of the paper is as follows.
In section~\ref{variant} we discuss the variant of the E$_6$SSM that we shall study,
and discuss the renormalisation group equations which constrain the
Yukawa couplings to be perturbative up to the unification scale.
In section~\ref{diphoton} we apply this model to the 750 GeV diphoton resonance,
calculating the branching ratios and cross-sections for the two scalar and two pseudoscalar
states associated with the $S_{1,2}$
singlets, including possible degeneracies and mixing.
Section~\ref{conclusions} concludes the paper.

\section{A variant of the E$_6$SSM}
\label{variant}
We first recall that the
E$_6$SSM~\cite{King:2005jy,King:2005my}
may be derived from an
$E_6$ GUT group broken via the following symmetry breaking chain:
\be
E_6 &\rightarrow& SO(10) \otimes U(1)_\psi \nn\\
&\rightarrow& SU(5) \otimes U(1)_\chi \otimes U(1)_\psi \nn\\
&\rightarrow& SU(3) \otimes SU(2) \otimes U(1)_Y
\times U(1)_\chi \otimes U(1)_\psi \nn\\
&\rightarrow& SU(3) \otimes SU(2) \otimes U(1)_Y \otimes U(1)_N.
\ee
We assume that the above symmetry breaking chain
occurs at a single GUT scale $M_X$ in one step,
due to some unspecified symmetry breaking sector,
\be
E_6 &\rightarrow& SU(3) \otimes SU(2) \otimes U(1)_Y \otimes U(1)_N,
\ee
where
\be
U(1)_N &=& \cos(\vartheta) U(1)_\chi + \sin(\vartheta) U(1)_\psi
\ee
and $\tan(\vartheta) = \sqrt{15}$ such that the right-handed neutrinos
that appear in the model are completely
neutral and may get large intermediate scale masses.
However the $U(1)_N$ gauge group remains unbroken down to the few TeV energy scale
where its breaking results in an observable $Z'_N$.
Three complete 27 representations of $E_6$ then also must survive
down to this scale in order to ensure anomaly cancellation. These $27_i$
decompose under the $SU(5) \otimes U(1)_N$ subgroup as follows:
\be
27_i &\rightarrow& (10,1)_i + (\bar{5},2)_i + (\bar{5},-3)_i + (5,-2)_i
+ (1,5)_i + (1,0)_i,
\ee
where the $U(1)_N$ charges must be GUT normalised by a factor of $1/\sqrt{40}$.
The first two terms contain the usual quarks and leptons, and the final term, which is
a singlet under the entire low energy gauge group, contains the
(CP conjugated) right-handed
neutrinos ${N}^c_i$. The last-but-one term, which is charged only under $U(1)_N$,
contains
the SM-singlet fields $S_i$.
The remaining terms $(\bar{5},-3)_i $ and $(5,-2)_i$
contain three families of
up- and down-type Higgs doublets, $H_{ui}$ and $H_{di}$,
and charged $\pm 1/3$ coloured exotics, $D_i$ and $\bar{D}_i$.
These are all superfields and are written with hats in the following.

The low energy gauge invariant superpotential can be written
\be
W^{\rr{E}_6\rr{SSM}} &=& W_0 + W_{1,2},\label{eq:w}
\ee
where $W_{0,1,2}$ are given by
\be
W_0 &=& \lambda_{jki}\hat{H}_{dj}\hat{H}_{uk}\hat{S}_i + \kappa_{jki}\hat{\bar{D}}_j\hat{D}_k \hat{S}_i + h^N_{ijk}\hat{N}^c_i\hat{H}_{uj}\hat{L}_{Lk}\label{eq:w0}\nn\\
&&\phantom{~} + h^U_{ijk}\hat{H}_{ui}\hat{Q}_{Lj}\hat{u}_{Rk}^c + h^D_{ijk}\hat{H}_{di}\hat{Q}_{Lj}\hat{d}_{Rk}^c + h^E_{ijk}\hat{H}_{di}\hat{L}_{Lj}\hat{e}_{Rk}^c,
\label{W0}\\
W_1 &=& g^Q_{ijk}\hat{D}_i\hat{Q}_{Lj}\hat{Q}_{Lk} + g^q_{ijk}\hat{\bar{D}}_i\hat{d}_{Rj}^c\hat{u}_{Rk}^c,\label{eq:w1}\\
W_2 &=& g^N_{ijk}\hat{N}^c_i\hat{D}_j\hat{d}_{Rk}^c + g^E_{ijk}\hat{D}_i\hat{u}_{Rj}^c\hat{e}_{Rk}^c +
 g^D_{ijk}\hat{\bar{D}}_i\hat{Q}_{Lj}\hat{L}_{Lk},\label{eq:w2}
\ee
with $W_{1,2}$ referring to either $W_1$ or $W_2$ (but not both together which would result in excessive
proton decay unless the associated Yukawa couplings were very small).

At the renormalisable level the gauge invariance ensures
matter parity and hence LSP stability. All lepton and quark superfields are defined to be odd under
matter parity $\dd{Z}_2^M$, while $\hat{H}_{ui}$, $\hat{H}_{di}$,  $\hat{D}_i$,
$\hat{\bar{D}}_i$, and $\hat{S}_i$ are even.
This means that the fermions associated with $\hat{D}_i$, $\hat{\bar{D}}_i$ are
SUSY particles analogous to the Higgsinos, while their scalar components may be thought of as colour-triplet
(and electroweak singlet) Higgses, making complete $5$ and $\bar{5}$ representations
without the usual doublet-triplet splitting.

In order for baryon and
lepton number to also be conserved, preventing rapid proton decay mediated by
$\hat{D}_i$, $\hat{\bar{D}}_i$, one imposes either
$\dd{Z}_2^{qq}$ or $\dd{Z}_2^{lq}$.
Under $\dd{Z}_2^{qq}$, the lepton, including the RH neutrino, superfields are assumed to be odd,
which forbids $W_2$.
Under $\dd{Z}_2^{lq}$,
the lepton and the $\hat{D}_i$, $\hat{\bar{D}}_i$ superfields are assumed odd,
which forbids $W_1$. Baryon and lepton number are conserved
at the renormalisable level,
with the $\hat{D}_i$, $\hat{\bar{D}}_i$ interpreted as being either diquarks in the former case or leptoquarks in the latter case.

In the E$_6$SSM,
a further approximate flavour symmetry $\dd{Z}_2^H$ was also assumed. It is this approximate
symmetry that distinguishes the third (by definition, ``active'') generation of Higgs doublets and
SM-singlets from the second and first (``inert'') generations. Under this approximate symmetry,
all superfields are
taken to be odd, apart from the active
$\hat{S}_3$, $\hat{H}_{d3}$, and $\hat{H}_{u3}$
which are taken to be even. The
inert fields then have small couplings to matter and do not radiatively acquire VEVs or
lead to large flavour changing neutral currents. The active fields can have large couplings to
matter and radiative electroweak symmetry breaking (EWSB) occurs with these fields.
In particular the VEV $\langle S_3 \rangle = s/\sqrt{2}$
is responsible for breaking the $U(1)_N$ gauge group and generating the
effective $\mu$ term and D-fermion masses. In particular we must have $s>5$~TeV in order to satisfy
$M_{Z'_N}>2.5$~TeV, which is the current LHC Run 2
experimental limit~\cite{CMS-1}.

We now propose a variant of
the E$_6$SSM in which we allow all three singlets $\hat{S}_i$
(as well as $\hat{H}_{d3}$ and $\hat{H}_{u3}$ )
to be even under the $\dd{Z}_2^H$.
This allows all three singlets $\hat{S}_i$ to
couple to $\hat{H}_{ui}$ and $\hat{H}_{di}$ as well as $\hat{D}_i$ and $\hat{\bar{D}}_i$.
If for simplicity we take the couplings in
Eq.~(\ref{W0}) to have the diagonal form,
$\lambda_{jki}\propto \lambda_{ji}\delta_{jk}$ and
$\kappa_{jki}\propto \kappa_{ji}\delta_{jk}$,
then the $Z^{H}_2$ symmetry allows to reduce the structure
of the Yukawa interactions in the superpotential to:
\be
W^{\rr{E}_6\rr{SSM}}\simeq \lambda_{ji}\hat{H}_{dj}\hat{H}_{uj}\hat{S}_i + \kappa_{ji}\hat{\bar{D}}_j\hat{D}_j \hat{S}_i
+ W_{MSSM}(\mu=0)\,.
\label{21}
\ee
The superfield $\hat{S}_3$ is assumed to
acquire a rather large VEV ($\langle S_3 \rangle = s/\sqrt{2}$) giving rise to the
effective $\mu$ term, masses of exotic quarks and inert Higgsino states which are given by
$$
\mu=\ds\frac{\lambda_{33} s}{\sqrt{2}}\,,\qquad\qquad \mu_{H_{\alpha}}=\ds\frac{\lambda_{\alpha 3} s}{\sqrt{2}}\,,
\qquad\qquad \mu_{D_{i}}=\ds\frac{\kappa_{i 3} s}{\sqrt{2}}\,,
$$
In our analysis here we restrict our consideration to the case when exotic quarks and inert Higgsinos are sufficiently light compared to
the VEV $s>5$ TeV, but are heavier than half the mass of the 750 GeV resonance, so that they appear in loop diagrams for the singlet decays.
It means that the Yukawa couplings of $\hat{S}_3$ to all exotic states should be quite small. Throughout this paper we are
going to assume that some scalar components of the first and second singlets $\hat{S}_{\alpha}$, with $\alpha = 1,2$,
can be identified with the resonances which give rise to the excess of diphoton events at an invariant mass around
$750\,\mbox{GeV}$ recently reported by the LHC experiments. ATLAS and CMS measurements indicate that the branching
ratios of the decays of such resonances into SM fermions have to be sufficiently small. This implies that the mixing between
the scalar components of $\hat{S}_{\alpha}$ and the neutral scalar components of the third pair of Higgs doublets $H_{u}$
and $H_{d}$, which are the ones that give rise to the EWSB, should be strongly suppressed. In order to ensure the suppression
of the corresponding mixing we impose the further requirement, namely that the SM singlets $\hat{S}_{\alpha}$, with $\alpha = 1,2$,
have rather small couplings to the third pair of Higgs doublets $H_{u}$ and $H_{d}$, i.e. $\lambda_{3\alpha}\approx 0$. This guarantees
that $\hat{S}_{\alpha}$ develop rather small VEVs and the mixing between the neutral scalar components of $\hat{S}_{\alpha}$,
$H_{u}$ and $H_{d}$ can be negligibly small so that it can be even ignored in the leading approximation. In this context it is
worth pointing out that if couplings $\kappa_{i3}$, $\lambda_{3\alpha}$, $\lambda_{\alpha 3}$ and $\lambda_{33}$ are set to
be small at the scale $M_X$ then they will remain small at any scale below $M_X$.

Neglecting the Yukawa couplings $\lambda_{3\alpha}$ the low energy effective superpotential of the modified E$_6$SSM
below the scale $\langle \hat{S}_3 \rangle$ can be written as
\begin{equation}
\begin{array}{c}
W_{eff} \simeq \lambda_{\alpha 1} \hat{S}_1 (\hat{H}^d_{\alpha} \hat{H}^u_{\alpha})+
\kappa_{i1} \hat{S}_1 (\hat{D}_i \hat{\bar{D}}_i)+\lambda_{\alpha 2} \hat{S}_2 (\hat{H}^d_{\alpha} \hat{H}^u_{\alpha})+
\kappa_{i2} \hat{S}_2 (\hat{D}_i \hat{\bar{D}}_i)\\[2mm]
+ \mu_{H_{\alpha}} (\hat{H}^d_{\alpha} \hat{H}^u_{\alpha}) + \mu_{D_i} (\hat{D}_i \hat{\bar{D}}_i) + W_{MSSM}(\mu \neq 0)\,.
\end{array}
\label{31}
\end{equation}
where $\alpha=1,2$ and $i=1,2,3$. The superpotential (\ref{31}) does not contain any
mass terms that involve superfields $\hat{S}_{\alpha}$. This implies that the fermion components of $\hat{S}_{\alpha}$
can be very light. In particular, the corresponding states can be lighter than $0.1\,\mbox{eV}$ forming hot
dark matter in the Universe. Such fermion states have negligible couplings to $Z$ boson as well as other
SM particles and therefore would not have been observed at earlier collider experiments. These states also
do not change the branching ratios of the $Z$ boson and Higgs decays\footnote{The presence of very
light neutral fermions in the particle spectrum might have interesting implications for the neutrino physics
(see, for example \cite{Frere:1996gb}).}. Moreover if $Z'$ boson is sufficiently
heavy the presence of such light fermion states does not affect Big Bang Nucleosynthesis \cite{Hall:2011zq}.

The superpotential (\ref{31})
contains ten new Yukawa couplings $\lambda_{\alpha 1}$, $\lambda_{\alpha 2}$, $\kappa_{i1}$ and $\kappa_{i2}$.
The running of these Yukawa couplings obey the following system of the renormalization group (RG) equations:
$$
\begin{array}{rcl}
\ds\frac{d \lambda_{\alpha\,1}}{dt} &=&\ds\frac{\lambda_{\alpha\,1}}{(4\pi)^2}\biggl[2\lambda_{\alpha\,1}^2+2\lambda_{\alpha\,2}^2+
2 (\sum_{\beta}\lambda^2_{\beta\,1}) + 3(\sum_j \kappa_{j1}^2) -3g_2^2\\[3mm]
&-&\ds\frac{3}{5}g_1^2-\ds\frac{19}{10} g^{'2}_1\biggr]
+\ds\frac{\lambda_{\alpha\,2}}{(4\pi)^2}\biggl[2 (\sum_{\beta}\lambda_{\beta\,1}\lambda_{\beta\,2}) + 3(\sum_j \kappa_{j1} \kappa_{j2}) \biggr]\,,\\[3mm]
\ds\frac{d \lambda_{\alpha\,2}}{dt} &=&\ds\frac{\lambda_{\alpha\,2}}{(4\pi)^2}\biggl[2\lambda_{\alpha\,1}^2+2\lambda_{\alpha\,2}^2+
2 (\sum_{\beta}\lambda^2_{\beta\,2}) + 3(\sum_j \kappa_{j2}^2) -3g_2^2\\[3mm]
&-&\ds\frac{3}{5}g_1^2-\ds\frac{19}{10} g^{'2}_1\biggr]
+\ds\frac{\lambda_{\alpha\,1}}{(4\pi)^2}\biggl[2 (\sum_{\beta}\lambda_{\beta\,1}\lambda_{\beta\,2}) + 3(\sum_j \kappa_{j1} \kappa_{j2}) \biggr]\,,
\end{array}
$$
\begin{equation}
\begin{array}{rcl}
\ds\frac{d \kappa_{i\,1}}{dt}&=&\ds\frac{\kappa_{i\,1}}{(4\pi)^2}\biggl[2\kappa^2_{i\,1}+2\kappa^2_{i\,2}+ 2 (\sum_{\beta}\lambda^2_{\beta\,1}) +
3(\sum_j \kappa_{j1}^2) - \ds\frac{16}{3}g_3^2\\[3mm]
&-&\ds\frac{4}{15}g_1^2-\ds\frac{19}{10} g^{'2}_1\biggr] +\ds\frac{\kappa_{i\,2}}{(4\pi)^2}\biggl[2 (\sum_{\beta}\lambda_{\beta\,1}\lambda_{\beta\,2})+
3(\sum_j \kappa_{j1} \kappa_{j2})\biggr]\,,\\[3mm]
\ds\frac{d \kappa_{i\,2}}{dt}&=&\ds\frac{\kappa_{i\,2}}{(4\pi)^2}\biggl[2\kappa^2_{i\,1}+2\kappa^2_{i\,2}+ 2 (\sum_{\beta}\lambda^2_{\beta\,2}) +
3(\sum_j \kappa_{j2}^2) - \ds\frac{16}{3}g_3^2\\[3mm]
&-&\ds\frac{4}{15}g_1^2-\ds\frac{19}{10} g^{'2}_1\biggr] +\ds\frac{\kappa_{i\,1}}{(4\pi)^2}\biggl[2 (\sum_{\beta}\lambda_{\beta\,1}\lambda_{\beta\,2})+
3(\sum_j \kappa_{j1} \kappa_{j2})\biggr]\,.
\end{array}
\label{32}
\end{equation}
The requirement of validity of perturbation theory up to the Grand Unification scale $M_X$ restricts the interval of variations of these Yukawa couplings
at low-energies. In our analysis here we use a set of one--loop RG equations (\ref{32}) while the evolution of gauge couplings is calculated
in the two--loop approximation.

\section{750 GeV diphoton excess in the variant E$_6$SSM}
\label{diphoton}

Turning now to a discussion of the $750\,\mbox{GeV}$ diphoton excess recently observed by ATLAS and
CMS in the framework of the variant of the E$_6$SSM discussed in the previous section,
whose effective superpotential is given by Eq.~(\ref{31}).
This SUSY model involves two SM singlet superfields $\hat{S}_{1,2}$
plus a set of extra
vector-like supermultiplets beyond the MSSM, including
two pairs of inert Higgs doublets ($\hat{H}^d_{\alpha}$ and
$\hat{H}^u_{\alpha}$), as well as three generations of exotic quarks $\hat{D}_i$
and $\overline{\hat{D}}_i$ with electric charges $\mp 1/3$.

The scenario discussed in this section
is that the 750-760 GeV diphoton resonance may be identified as one or two scalars denoted
$N_{1,2}$ and/or
one or two pseudoscalars denoted $A_{1,2}$
contained in the two singlet superfields $\hat{S}_{1,2}$.
The masses of these scalars and pseudoscalars arises from the soft SUSY breaking sector.
However, to simplify our analysis, we assume that all other
sparticles are sufficiently heavy so that their contributions
to the production and decay rates of states with masses around $750\,\mbox{GeV}$ can be ignored. Moreover
the scenario under consideration implies that almost all exotic vector-like fermion mass
states are heavier than $375\,\mbox{GeV}$ so
that the on-shell decays of $N_{\alpha}$ and $A_{\alpha}$ into the corresponding particles are not kinematically allowed.

Integrating out the heavy
fermions corresponding to two pairs of inert Higgsino doublets
$\tilde{H}^d_{\alpha}$ and
$\tilde{H}^u_{\alpha}$
and three generations of vector-like ${D}_i$
and $\overline{{D}}_i$ fermions,
which appear in the usual triangle loop diagrams,
one obtains the effective Lagrangian which describes the interactions of
$N_{\alpha}$ and $A_{\alpha}$  with the SM gauge bosons,
\begin{equation}
\begin{array}{c}
\mathcal{L}_{eff}=\sum_{\alpha}\Biggl( c_{1\alpha} N_{\alpha} B_{\mu\nu} B^{\mu\nu} + c_{2\alpha} N_{\alpha} W^a_{\mu\nu} W^{a\mu\nu}
+ c_{3\alpha} N_{\alpha} G^{\sigma}_{\mu\nu} G^{\sigma\mu\nu} \qquad\qquad\qquad\qquad\\[0mm]
\qquad\qquad\qquad\qquad\qquad\qquad + \tilde{c}_{1\alpha} A_{\alpha} B_{\mu\nu} \widetilde{B}^{\mu\nu} +
\tilde{c}_{2\alpha} A_{\alpha} W^a_{\mu\nu} \widetilde{W}^{a\mu\nu} + \tilde{c}_{3\alpha} A_{\alpha} G^{\sigma}_{\mu\nu} \widetilde{G}^{\sigma\mu\nu}
\Biggr)\,,
\end{array}
\label{33}
\end{equation}
where
\begin{equation}
\begin{array}{rcl}
c_{1\alpha}&=&\ds\frac{\alpha_Y}{16\pi}\Biggl[\sum_i \frac{2\kappa_{i\alpha}}{3\sqrt{2}\mu_{D_i}} A(x_{D_i}) +
\sum_{\beta} \frac{\lambda_{\beta\alpha}}{\sqrt{2}\mu_{H_{\beta}}} A(x_{H_{\beta}})\Biggr]\,,\\
c_{2\alpha}&=&\ds\frac{\alpha_2}{16\pi}\Biggl[\sum_{\beta} \frac{\lambda_{\beta\alpha}}{\sqrt{2}\mu_{H_{\beta}}} A(x_{H_{\beta}}) \Biggr]\,,\\
c_{3\alpha}&=&\ds\frac{\alpha_3}{16\pi}\Biggl[\sum_i \frac{\kappa_{i\alpha}}{\sqrt{2}\mu_{D_i}} A(x_{D_i})\Biggr]\,,\\[6mm]
A(x) &=& 2 x (1 + (1-x)\arcsin^2[1/\sqrt{x}])\,,\qquad \mbox{for} \qquad x\ge 1\,.
\end{array}
\label{34}
\end{equation}
In Eq. (\ref{33}) $B_{\mu\nu}$, $W^a_{\mu\nu}$, $G^{\sigma}_{\mu\nu}$ are field strengths for the $U(1)_Y$, $SU(2)_W$ and $SU(3)_C$
gauge interactions respectively while $\widetilde{G}^{\sigma\mu\nu}=\frac{1}{2}\epsilon^{\mu\nu\lambda\rho} G^{\sigma}_{\lambda\rho}$ etc.
In Eqs.~(\ref{34}) $x_{D_i}=4\mu_{D_i}^2/M_X^2$, $x_{H_{\alpha}}=4\mu_{H_{\alpha}}^2/M_X^2$ and $\alpha_Y=3\alpha_1/5$
whereas $\alpha_1$, $\alpha_2$ and $\alpha_3$ are (GUT normalised) gauge couplings of $U(1)_Y$, $SU(2)_W$ and $SU(3)_C$ interactions.
In order to obtain analytic expressions for $\tilde{c}_{i\alpha}$ one should replace in Eqs.~(\ref{34}) $c_{i\alpha}$ by $\tilde{c}_{i\alpha}$
and substitute function $B(x)$ instead of $A(x)$, where
\begin{equation}
B(x)=2 x \arcsin^2[1/\sqrt{x}]\,.
\label{35}
\end{equation}

Because in our analysis we focus on the diphoton  decays of $N_{\alpha}$ and $A_{\alpha}$ that may lead to the 750 GeV diphoton excess
it is convenient to use the effective Lagrangian that describes the interactions of these fields with the electromagnetic one. Using Eq.~(\ref{33})
one obtains
\begin{equation}
\mathcal{L}^{\gamma\gamma}_{eff}=\sum_{\alpha} \Biggl( c^{\gamma}_{\alpha} N_{\alpha} F_{\mu\nu} F^{\mu\nu} +
\tilde{c}^{\gamma}_{\alpha} A_{\alpha} F_{\mu\nu} \widetilde{F}^{\mu\nu}\Biggr)\,,
\label{36}
\end{equation}
where $c^{\gamma}_{\alpha}=c_{1\alpha}\cos^2\theta_W+ c_{2\alpha}\sin^2\theta_W$,
$\tilde{c}^{\gamma}_{\alpha}=\tilde{c}_{1\alpha}\cos^2\theta_W+ \tilde{c}_{2\alpha}\sin^2\theta_W$ and $F_{\mu\nu}$ is
a field strength associated with the electromagnetic interaction.

At the LHC the exotic states $N_{\alpha}$ and $A_{\alpha}$ can be predominantly produced through gluon fusion. When exotic
quarks have masses below $1\,\mbox{TeV}$ the corresponding production cross section is rather large and determined by the
effective couplings $|c_{3\alpha}|^2$ and $|\tilde{c}_{3\alpha}|^2$. However such states mainly decay into a pair of gluons
which is very problematic to detect at the LHC. Therefore possible collider signatures of these exotic states are associated with
their decays into $WW$, $ZZ$, $\gamma Z$ and $\gamma\gamma$. Since $W$ and $Z$ decay mostly into quarks the process
$pp\to N_{\alpha}(A_{\alpha})\to \gamma\gamma$ tends to be one of the most promising channels to search for such resonances.
In the limit when exotic states decay predominantly into a pair of gluons the branching ratios of $N_{\alpha}\to \gamma\gamma$
and $A_{\alpha}\to \gamma\gamma$ are proportional to $|c^{\gamma}_{\alpha}|^2/|c_{3\alpha}|^2$ and
$|\tilde{c}^{\gamma}_{\alpha}|^2/|\tilde{c}_{3\alpha}|^2$ respectively. As a consequence cross sections
$\sigma(pp\to N_{\alpha}(A_{\alpha})\to \gamma\gamma)$ do not depend on $|c_{3\alpha}|^2$ and $|\tilde{c}_{3\alpha}|^2$.
The corresponding signal strengths are basically defined by the partial decay widths $\Gamma(N_{\alpha}\to \gamma\gamma)$
and $\Gamma(A_{\alpha}\to \gamma\gamma)$.

The cross sections of the processes that may result in the 750 GeV diphoton excess can be written as
\begin{equation}
\sigma(pp\to X_{\alpha} \to \gamma\gamma)\simeq \ds\frac{C_{gg}}{M_{X_{\alpha}} s \Gamma_{X_{\alpha}}} \Gamma(X_{\alpha}\to gg) \Gamma(X_{\alpha}\to \gamma\gamma)\,,
\label{37}
\end{equation}
where $X_{\alpha}$ is either $N_{\alpha}$ or $A_{\alpha}$ exotic states, $\Gamma_{X_{\alpha}}$ is a total decay width of the resonance $X_{\alpha}$
while $C_{gg}\simeq 3163$, $\sqrt{s}\simeq 13\,\mbox{TeV}$ and $M_{X_{\alpha}}$ is the mass of the appropriate exotic state which should be
somewhat around $750\,\mbox{GeV}$. The partial decay widths of the corresponding resonances are given by
\begin{equation}
\begin{array}{c}
\Gamma(N_{\alpha}\to gg)=\ds\frac{2}{\pi} M^3_{N_{\alpha}} |c_{3\alpha}|^2\,,\qquad \Gamma(A_{\alpha}\to gg)=\ds\frac{2}{\pi} M^3_{A_{\alpha}} |\tilde{c}_{3\alpha}|^2\,,\\[3mm]
\Gamma(N_{\alpha}\to \gamma\gamma)=\ds\frac{M^3_{N_{\alpha}}}{4\pi} |c^{\gamma}_{\alpha}|^2\,,\qquad
\Gamma(A_{\alpha}\to \gamma\gamma)=\ds\frac{M^3_{A_{\alpha}}}{4\pi} |\tilde{c}^{\gamma}_{\alpha}|^2\,.
\end{array}
\label{38}
\end{equation}
In the limit when $\Gamma_{X_{\alpha}}\approx \Gamma(X_{\alpha}\to gg)$ the dependence of the cross section (\ref{37}) on $\Gamma(X_{\alpha}\to gg)$
disappear and its value is determined by the partial decay width $\Gamma(X_{\alpha}\to \gamma\gamma)$ as one could naively expect.
In this case, as it was pointed out in \cite{Franceschini:2015kwy}, one can obtain $\sigma(pp\to \gamma\gamma)\approx 8\,\mbox{fb}$
at the $13\,\mbox{TeV}$ LHC if
\begin{equation}
\ds\frac{\Gamma(X_{\alpha}\to \gamma\gamma)}{M_{X_{\alpha}}}=1.1\times 10^{-6}\,.
\label{39}
\end{equation}
Then the cross section $\sigma_{\gamma\gamma}\approx \sigma(pp\to \gamma\gamma)$ for arbitrary partial decay widths of $X_{\alpha}\to \gamma\gamma$
can be approximately estimated as
\begin{equation}
\sigma_{\gamma\gamma}\simeq 7.3\, \mbox{fb}\times \mbox{BR}(X_{\alpha}\to gg)\times \left( \ds\frac{\Gamma(X_{\alpha}\to \gamma\gamma)}{M_{X_{\alpha}}}\times 10^{6}\right)\,.
\label{391}
\end{equation}
where the branching ratios associated with the decays of exotic states into gluons $g$ and vector bosons $V$ ($V=\gamma,\, W^{\pm}, Z$) are given by
\begin{equation}
\mbox{BR}(X_{\alpha}\to gg)=\frac{\Gamma(X_{\alpha}\to gg)}{\Gamma_{X_{\alpha}}}\,,\qquad
\mbox{BR}(X_{\alpha}\to VV)=\frac{\Gamma(X_{\alpha}\to VV)}{\Gamma_{X_{\alpha}}}\,.
\label{392}
\end{equation}
In Eqs.~(\ref{392}) $\Gamma(X_{\alpha}\to gg)$ and $\Gamma(X_{\alpha}\to VV)$ are partial decay widths that correspond to the exotic state decays
into a pair of gluons and a pair of vector bosons respectively whereas $\Gamma_{X_{\alpha}}$ is a total decay width of this state.

\subsection{One scalar/pseudoscalar case}

Let us now consider the scenario when one of the scalar/pseudoscalar exotic states ($N_1$ or $A_1$) has a mass which is
rather close to $750\,\mbox{GeV}$.  From Eqs.~(\ref{34})--(\ref{36}) and (\ref{38}) it follows that the diphoton decay rates
of these new bosons and the corresponding signal strength depend very strongly on the values of the Yukawa coulings
$\lambda_{\alpha 1}$ and $\kappa_{i 1}$. On the other hand the growth of these Yukawa couplings at low energies
entails the increase of their values at the Grand Unification scale $M_X$ resulting in the appearance of the Landau pole
that spoils the applicability of perturbation theory at high energies (see, for example \cite{Nevzorov:2001vj}).
The requirement of validity of perturbation theory up to the scale $M_X$ sets an upper bound on the low energy value of $\lambda_{\alpha 1}$
and $\kappa_i$. In our analysis we use two--loop SM RG equations to compute the values of the gauge couplings at the scale $Q=2\,\mbox{TeV}$.
Above this scale we use two--loop RG equations for the gauge couplings and one---loop RG equations for the Yukawa couplings
including the ones given by Eq.~(\ref{32}) to analyse the RG flow of these couplings. In the simplest case when $\lambda_{\alpha 1}=\kappa_{i 1}$
our numerical analysis indicates that the values of these couplings at the scale $Q=2\,\mbox{TeV}$ should not exceed $0.6$.

\begin{figure}
\hspace*{0cm}{$\mbox{BR}(A_{1} \to gg, WW, ZZ, \gamma\gamma, \gamma Z)$}
\hspace*{3cm}{$\mbox{BR}(N_{1} \to gg, WW, ZZ, \gamma\gamma, \gamma Z)$}\\[-4mm]
\includegraphics[width=75mm,height=55mm]{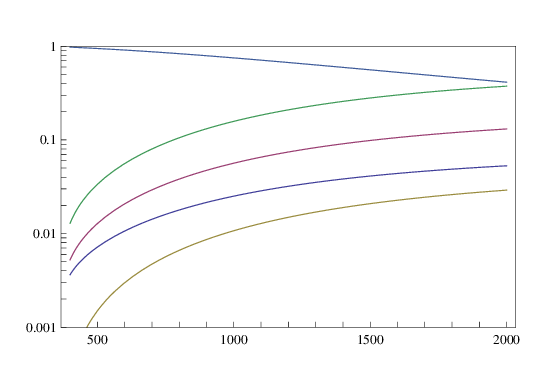}\qquad
\includegraphics[width=75mm,height=55mm]{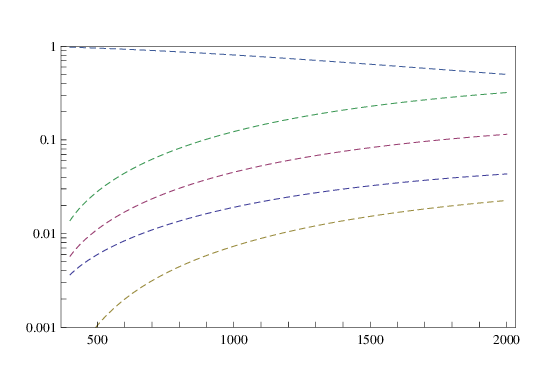}\\[-5mm]
\hspace*{3.5cm}{$\mu_D$}\hspace*{7.8cm}{$\mu_D$}\\[1mm]
\hspace*{3.5cm}{\bf (a)}\hspace*{7.5cm}{\bf (b) }\\[2mm]
\hspace*{0cm}{$\frac{\Gamma(A_{1} \to \gamma\gamma)}{M_X} \times 10^6$}\hspace*{6cm}{$\frac{\Gamma(N_{1} \to \gamma\gamma)}{M_X} \times 10^6$}\\[-4mm]
\includegraphics[width=75mm,height=55mm]{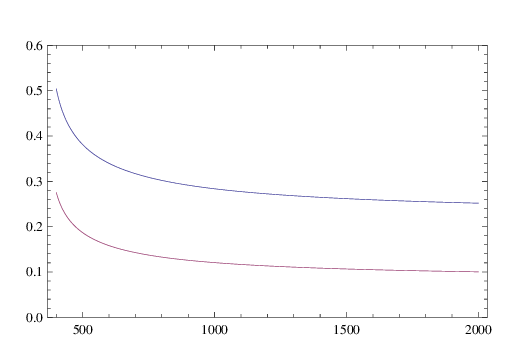}\qquad
\includegraphics[width=75mm,height=55mm]{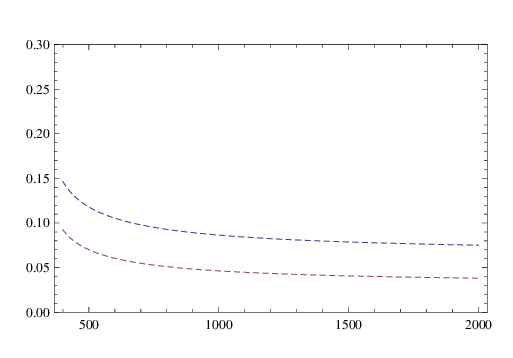}\\[-5mm]
\hspace*{3.5cm}{$\mu_D$}\hspace*{7.8cm}{$\mu_D$}\\[1mm]
\hspace*{3.5cm}{\bf (c)}\hspace*{7.5cm}{\bf (d) }\\[2mm]
\hspace*{0cm}{$\sigma(pp\to A_{1} \to \gamma\gamma)[\mbox{fb}]$}
\hspace*{5.0cm}{$\sigma(pp\to N_{1} \to \gamma\gamma)[\mbox{fb}]$}\\[-5mm]
\includegraphics[width=75mm,height=55mm]{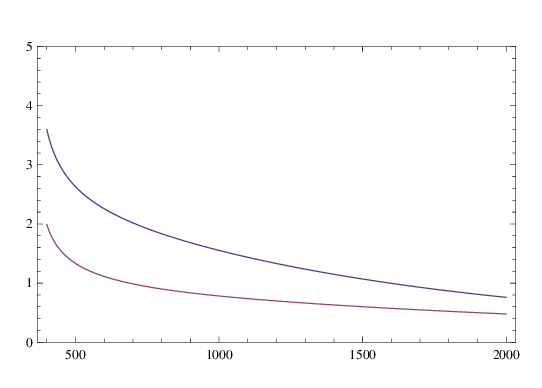}\qquad
\includegraphics[width=75mm,height=55mm]{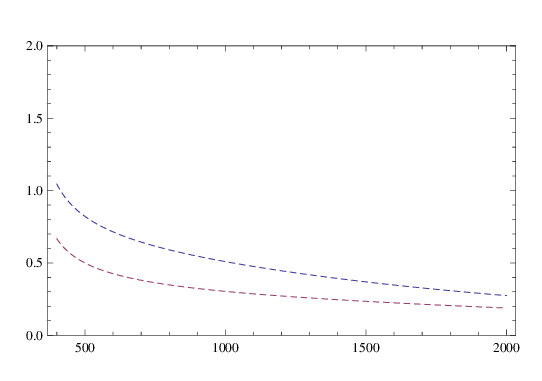}\\[-5mm]
\hspace*{3.5cm}{$\mu_D$}\hspace*{7.8cm}{$\mu_D$}\\[1mm]
\hspace*{3.5cm}{\bf (e)}\hspace*{7.5cm}{\bf (f)}
\caption{ Predictions for the one pseudoscalar (left panels) or one scalar (right panels) case.
In all cases the masses of vector-like quarks are set to be equal, i.e. $\mu_{D_i}=\mu_D$, whereas
$\lambda_{\alpha 1}=\kappa_{i1}=0.6$,
$\lambda_{\alpha 2}=\kappa_{i 2}=0$.
In {\it(a)} the branching ratios of the decays of $A_1$ into $\gamma Z$ (lowest solid line),
$\gamma\gamma$ (second lowest solid line), $ZZ$ (third lowest solid line), $WW$ (second highest solid line)
and $gg$ (highest solid line) as a function of exotic quark masses $\mu_D$ for $M_{A_1}\simeq 750\,\mbox{GeV}$.
In {\it (b)} the branching ratios of the decays of $N_1$ into $\gamma Z$ (lowest dashed line),
$\gamma\gamma$ (second lowest dashed line), $ZZ$ (third lowest dashed line), $WW$ (second highest dashed line)
and $gg$ (highest dashed line) as a function of $\mu_D$ for $M_{N_1}\simeq 750\,\mbox{GeV}$.
In {\it (c)} the ratios ${\Gamma(A_1\to \gamma\gamma)}/{M_X}$ as a function of $\mu_D$ for $M_{A_1}\simeq 750\,\mbox{GeV}$.
The upper and lower solid lines correspond to the scenarios with $\mu_{H_{\alpha}}=400\,\mbox{GeV}$ and $\mu_{H_{\alpha}}=500\,\mbox{GeV}$.
In {\it (d)} the ratios ${\Gamma(N_1\to \gamma\gamma)}/{M_X}$ as a function of $\mu_D$ for $M_{N_1}\simeq 750\,\mbox{GeV}$.
The upper and lower dashed lines correspond to the scenarios with $\mu_{H_{\alpha}}=400\,\mbox{GeV}$ and $\mu_{H_{\alpha}}=500\,\mbox{GeV}$.
In {\it (e)} the cross sections $\sigma(pp\to A_{1} \to \gamma\gamma)$ in fb
as a function of $\mu_D$ for $M_{A_1}\simeq 750\,\mbox{GeV}$.
The upper and lower solid lines represent the scenarios with $\mu_{H_{\alpha}}=400\,\mbox{GeV}$ and $\mu_{H_{\alpha}}=500\,\mbox{GeV}$.
In {\it (f)} the cross sections $\sigma(pp\to N_{1} \to \gamma\gamma)$ in fb as a function of $\mu_D$ for $M_{N_1}\simeq 750\,\mbox{GeV}$.
The upper and lower dashed lines represent the scenarios with $\mu_{H_{\alpha}}=400\,\mbox{GeV}$ and $\mu_{H_{\alpha}}=500\,\mbox{GeV}$.
}
\label{fig1}
\end{figure}

The upper bound on the coupling $\lambda_{\alpha 1}$ becomes less stringent when $\kappa_{i 1}$ are small. In the limit when
all $\kappa_{i 1}$ vanish the value of $\lambda_{1 1}=\lambda_{2 1}$ has to remain smaller than $0.81$ to ensure the applicability
of perturbation theory up to the GUT scale. Although in this case  $\Gamma(A_1\to \gamma\gamma)$ and $\Gamma(N_1\to \gamma\gamma)$
attain their maximal value the production cross sections of exotic states $N_1$ or $A_1$ are negligibly small since
they are determined by the low--energy values of $\kappa_{i 1}$. The upper bounds on $\kappa_{i 1}$ can be also significantly relaxed
when $\lambda_{1 1}=\lambda_{2 1}=0$. If this is a case then the requirement of the validity of perturbation theory implies that
$\kappa_{1 1}=\kappa_{2 1}=\kappa_{3 1}\lesssim 0.79$. However in this limit the diphoton production rate associated with the
presence $A_1$ or $N_1$ is again negligibly small because the corresponding partial decay width vanish. Thus in this section
we focus on the scenario with $\lambda_{\alpha 1}=\kappa_{i 1}=0.6$. This choice of parameters guarantees that the production
cross sections of $N_1$ and $A_1$ as well as their partial decay width can be sufficiently large.

In Figs.~1a and 1b the dependence of the branching ratios of the exotic pseudoscalar and scalar states on the masses of exotic quarks
is examined. To simplify our analysis the masses of all exotic quarks are set to be equal while the masses of all inert Higgsinos
are assumed to be around $400\,\mbox{GeV}$. From Fig.~1a and 1b it follows that the exotic pseudoscalar and scalar states decay
predominantly into a pair of gluons when the masses of exotic quarks $\mu_D$ are below $1\,\mbox{TeV}$. Moreover if $\mu_D$
is close to $400-500\,\mbox{GeV}$ all other branching ratios are negligibly small. With increasing $\mu_D$ the branching ratio
of the exotic pseudoscalar (scalar) state decays into gluons decreases whereas the branching ratios of the decays of this state
into $W^{+}W^{-}$, $ZZ$, $\gamma\gamma$ and $\gamma Z$ increase. The branching ratios of $A_{1}(N_1)\to WW$ and
$A_{1}(N_1)\to ZZ$ are the second and third largest ones. The branching ratio of  $A_{1}(N_1)\to \gamma\gamma$ is considerably
smaller but still larger than $A_{1}(N_1)\to \gamma Z$. Although the branching ratios of $A_{1}(N_1)\to WW$ and
$A_{1}(N_1)\to ZZ$ can be a substantially bigger than the branching ratio  $A_{1}(N_1)\to \gamma\gamma$ their experimental
detection is more problematic because $W$ and $Z$ decays mainly into quarks. When $\mu_D$ is around $1\,\mbox{TeV}$
the branching ratio of $A_{1}(N_1)\to gg$ is still the largest one and constitutes about $75\% (80\%)$ while for $\mu_D\simeq 2\,\mbox{TeV}$
the branching ratios of $A_{1}(N_1)\to gg$ and $A_{1}(N_1)\to WW$ become comparable.

In Fig.~1c and 1d we explore the dependence of the partial decay widths associated with the decays of the exotic pseudoscalar and scalar states
into a pair of photons on the masses of exotic quarks and inert Higgsinos $\mu_{H_{\alpha}}$. One can see that these decay widths decrease
very rapidly with increasing $\mu_{H_{\alpha}}$. The dependence on the masses of exotic quarks is weaker because these states carry small
electric charges $\pm 1/3$. Since here we assume that $\kappa_{i 1}/\mu_{D_i}$ and $\lambda_{\alpha 1}/\mu_{H_{\alpha}}$ have the same
sign the growth of either exotic quark masses or $\mu_{H_{\alpha}}$ results in the reduction of the corresponding decay rate. When $\mu_D$
is larger than $1.5\,\mbox{TeV}$ the dependence of the partial decay widths under consideration becomes rather weak. From Fig.~1c and 1d
it is easy to see that the partial width of the decays $A_1\to \gamma\gamma$ is substantially larger than the one associated with
$N_1\to \gamma\gamma$ leading to the larger value of the cross sections $\sigma(pp \to A_{1} \to \gamma\gamma)$ as compared with
$\sigma(pp \to N_{1} \to \gamma\gamma)$.

In our analysis we use Eq.~(\ref{391}) to estimate the values of the cross sections $\sigma(pp \to A_{1} \to \gamma\gamma)$ and
$\sigma(pp \to N_{1} \to \gamma\gamma)$ at the $13\,\mbox{TeV}$ LHC. The results of our investigation are shown in Figs.~1e and 1f.
In the case of scalar exotic states with mass $750\,\mbox{GeV}$ this cross section tends to be substantially smaller than $1\,\mbox{ fb}$.
The presence of $750\,\mbox{GeV}$ exotic pseudoscalar can lead to the considerably stronger signal in the diphoton channel. When all
exotic quarks have masses around $400-500\,\mbox{GeV}$ the corresponding cross section can reach $2-3\,\mbox{fb}$. Somewhat stronger
signal can be obtained if we assume that both scalar and pseudoscalar exotic states have masses which are close to $750\,\mbox{GeV}$.
In this case the sum of the cross sections $\sigma(pp \to A_{1} \to \gamma\gamma)+\sigma(pp \to N_{1} \to \gamma\gamma)$ can
reach $4.5\,\mbox{fb}$ if exotic quarks have masses about $400\,\mbox{GeV}$. The existence of two nearly degenerate resonances
may also explain why the analysis performed by the ATLAS collaboration leads to the relatively large best-fit width which is about $45\,\mbox{GeV}$.
Unfortunately, the cross sections mentioned above decreases substantially with increasing exotic quark masses. Indeed, if $\mu_D\gtrsim 1\,\mbox{TeV}$
the sum of the cross sections $\sigma(pp \to A_{1} \to \gamma\gamma)+\sigma(pp \to N_{1} \to \gamma\gamma)$ does not exceed $2\,\mbox{fb}$.
These cross sections continue to fall even for $\mu_D\gtrsim 1.5\,\mbox{TeV}$ when the corresponding partial decay widths are rather close to their
lower saturation limits because the branching ratios associated with the decays of $A_1$ and $N_1$ into a pair of gluons decrease with
increasing $\mu_D$.

\begin{figure}
\hspace*{0cm}{$\mbox{BR}(A_{1,2} \to gg, WW, ZZ, \gamma\gamma, \gamma Z)$}
\hspace*{3cm}{$\mbox{BR}(N_{1,2} \to gg, WW, ZZ, \gamma\gamma, \gamma Z)$}\\[-4mm]
\includegraphics[width=75mm,height=55mm]{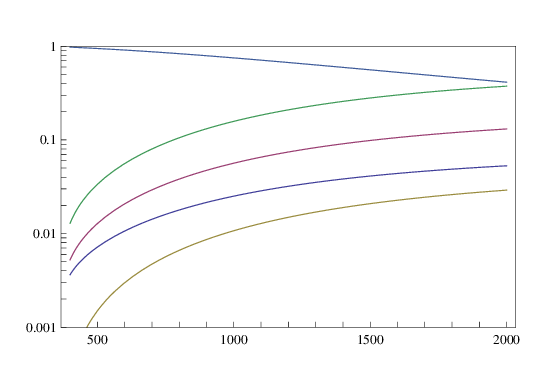}\qquad
\includegraphics[width=75mm,height=55mm]{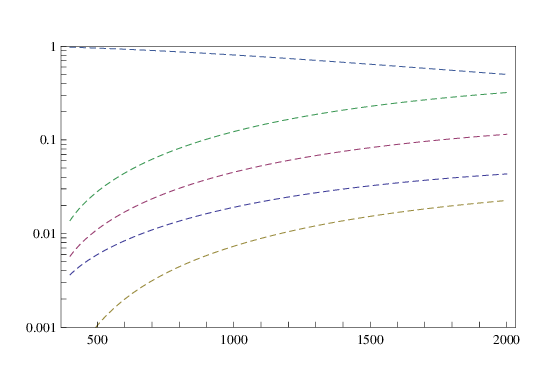}\\[-5mm]
\hspace*{3.5cm}{$\mu_D$}\hspace*{7.8cm}{$\mu_D$}\\[1mm]
\hspace*{3.5cm}{\bf (a)}\hspace*{7.5cm}{\bf (b) }\\[2mm]
\hspace*{0cm}{$\frac{\Gamma(A_{1,2}\to \gamma\gamma)}{M_X}\times 10^6$}
\hspace*{5.5cm}{$\frac{\Gamma(N_{1,2}\to \gamma\gamma)}{M_X}\times 10^6$}\\[-4mm]
\includegraphics[width=75mm,height=55mm]{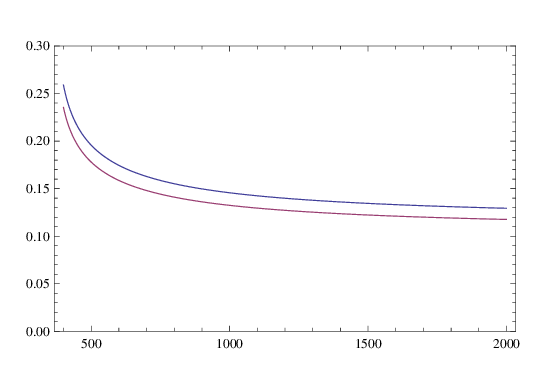}\qquad
\includegraphics[width=75mm,height=55mm]{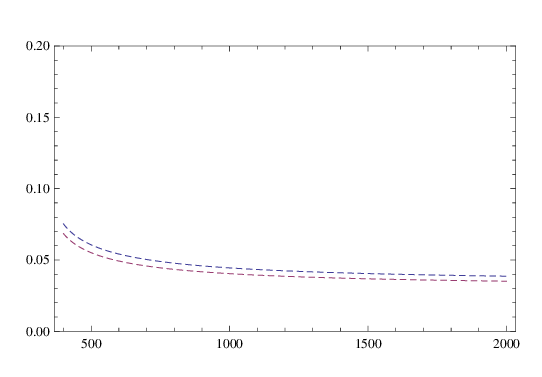}\\[-5mm]
\hspace*{3.5cm}{$\mu_D$}\hspace*{7.8cm}{$\mu_D$}\\[1mm]
\hspace*{3.5cm}{\bf (c)}\hspace*{7.5cm}{\bf (d) }\\[2mm]
\hspace*{0cm}{$\sigma(pp\to A_{1,2} \to \gamma\gamma)[\mbox{fb}]$}
\hspace*{4.7cm}{$\sigma(pp\to N_{1,2} \to \gamma\gamma)[\mbox{fb}]$}\\[-5mm]
\includegraphics[width=75mm,height=55mm]{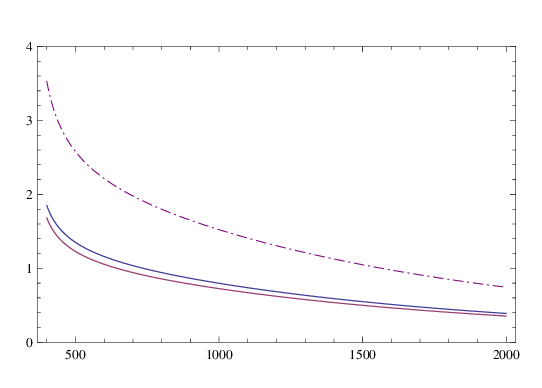}\qquad
\includegraphics[width=75mm,height=55mm]{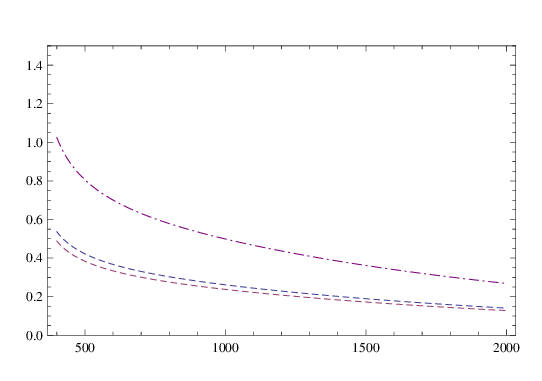}\\[-5mm]
\hspace*{3.5cm}{$\mu_D$}\hspace*{7.8cm}{$\mu_D$}\\[1mm]
\hspace*{3.5cm}{\bf (e)}\hspace*{7.5cm}{\bf (f)}
\caption{Predictions for two degenerate pseudoscalars (left panels) or two degenerate
scalars (right panels) case.
In all cases $\mu_{D_i}=\mu_D$, while $\mu_{H_{\alpha}}=400\,\mbox{GeV}$,
$\lambda_{\alpha 1}=\kappa_{i1}=0.43$ and $\lambda_{\alpha 2}=\kappa_{i2}=0.41$.
In {\it (a)} the branching ratios of the decays of $A_{1,2}$ into $\gamma Z$ (lowest solid line),
$\gamma\gamma$ (second lowest solid line), $ZZ$ (third lowest solid line), $WW$ (second highest solid line)
and $gg$ (highest solid line) as a function of exotic quark masses $\mu_D$ for $M_{A_{1,2}}\simeq 750\,\mbox{GeV}$.
In {\it (b)} the branching ratios of the decays of $N_{1,2}$ into $\gamma Z$ (lowest dashed line),
$\gamma\gamma$ (second lowest dashed line), $ZZ$ (third lowest dashed line), $WW$ (second highest dashed line)
and $gg$ (highest dashed line) as a function of $\mu_D$ for $M_{N_{1,2}}\simeq 750\,\mbox{GeV}$.
In {\it (c)} the ratios ${\Gamma(A_{1}\to \gamma\gamma)}/{M_X}$ (upper solid line) and
${\Gamma(A_{2}\to \gamma\gamma)}/{M_X}$ (lower solid line)
as a function of $\mu_D$ for $M_{A_{1,2}}\simeq 750\,\mbox{GeV}$.
In {\it (d)} the ratios $\Gamma(N_{1}\to \gamma\gamma)/M_X$ (upper dashed line)
and $\Gamma(N_{2}\to \gamma\gamma)/M_X$ (lower dashed line) as a function of $\mu_D$ for $M_{N_{1,2}}\simeq 750\,\mbox{GeV}$.
In {\it (e)} the cross sections (fb) $\sigma(pp\to A_{1} \to \gamma\gamma)$ (upper solid line) and $\sigma(pp\to A_{2} \to \gamma\gamma)$ (lower solid line) as a function
of $\mu_D$ for $M_{A_{1,2}}\simeq 750\,\mbox{GeV}$. The dashed--dotted line correspond to the sum of these cross sections.
In {\it (f)} the cross sections (fb) $\sigma(pp\to N_{1} \to \gamma\gamma)$ (upper dashed line) and
$\sigma(pp\to N_{2} \to \gamma\gamma)$ (lower dashed line)
as a function of $\mu_D$ for $M_{N_{1,2}}\simeq 750\,\mbox{GeV}$. The dashed--dotted line correspond to the sum of these cross sections.
}
\label{fig2}
\end{figure}

\subsection{Two degenerate scalar/pseudoscalar case}

Now let us assume that there are two superfields $\hat{S}_1$ and $\hat{S}_2$ that have sufficiently large Yukawa couplings to
the exotic quark and inert Higgsino states and can contribute to the measured cross section $pp\to \gamma\gamma$.
In other words we assume that scalar and pseudoscalar components of both superfields can have masses around
$750\,\mbox{GeV}$. Naively one may expect that this could allow to enhance the theoretical prediction for the
cross section $pp\to \gamma\gamma$. Again we start from the simplest case when all Yukawa couplings are
the same. Then the numerical analysis indicates that in this case the requirement of the validity of perturbation
theory up to the scale $M_X$ sets even more stringent upper bound on the low energy value of the Yukawa couplings
as compared with the one scalar/pseudoscalar case. Indeed, using the one--loop RG equations (\ref{32}) and two--loop
RG equations for the gauge couplings one obtains that $\lambda_{\alpha 1}=\kappa_{i 1}=\lambda_{\alpha 2}=\kappa_{i 2}=\lambda_0 \lesssim 0.43$.
Smaller values of the Yukawa couplings do not affect the branching ratios of $A_1$ and $N_1$. Moreover $A_2$ and $A_1$
as well as $N_2$ and $N_1$ have basically the same branching ratios. This is because partial decay widths of $A_{1,2}$ and
$N_{1,2}$ as well as the corresponding total widths are proportional to $\lambda_0^2$. As a consequence in the
leading approximation branching ratios do not depend on $\lambda_0$ (see Fig.~2a and 2b ). On the other hand
as one can see from Fig.~1c,\,1d,\,1e and 1f the partial decay widths of $A_{1,2}\to \gamma\gamma$ and
$N_{1,2}\to \gamma\gamma$ as well as the cross sections $\sigma(pp\to A_{1,2} \to \gamma\gamma)$
and $\sigma(pp\to N_{1,2} \to \gamma\gamma)$ are reduced by factor $2$ because of the smaller values of
the Yukawa couplings. If all exotic states $A_1$ and  $A_2$ as well as $N_1$ and $N_2$ are nearly degenerate
around $750\,\mbox{GeV}$ so that their distinction is not possible within present experimental accuracy,
then the superpositions of rates from these bosons basically reproduces the corresponding rates in the
one scalar/pseudoscalar case (see Figs.~1e, 1f, 2e and 2f). Thus, it seems rather problematic to achieve any enhancement
of the signal in the diphoton channel in the scenario when all Yukawa couplings are equal or reasonably close to each other.

\subsection{Maximal mixing scenario}

Following on from the discussion in the previous subsection,
there is one case when a modest enhancement of the signal in the diphoton channel
can be achieved. This happens in the so--called maximal mixing scenario when the masses of exotic scalars
as well as the masses of exotic pseudoscalars are rather close to $750\,\mbox{GeV}$ and the breakdown of SUSY gives rise to the
mixing of these states preserving CP conservation. In this case one can expect that the mixing angles between CP--odd exotic
states and CP--even exotic states tend to be rather large, i.e.about $\pm \pi/4$, because these bosons are nearly degenerate.
To simplify our analysis here we set these angles to be equal to $\pi/4$. Then the scalar components of the superfields
$S_1$ and $S_2$ can be expressed in terms of the mass eigenstates $N_1$, $N_2$, $A_1$ and $A_2$ as follows
\begin{equation}
S_1=\ds\frac{1}{2}\left(N_1 + N_2 + i(A_1+A_2) \right)\,,\qquad\qquad
S_2=\ds\frac{1}{2}\left(N_1 - N_2 + i(A_1-A_2) \right)\,.
\label{393}
\end{equation}
In addition we assume that only superfield $S_1$ couples to the inert Higgsino states, i.e. $\lambda_{\alpha 2}=0$, and
only superfield $S_2$ couples to the exotic quarks, i.e. $\kappa_{i 1}=0$. In this limit the requirement of the validity of perturbation
theory up to the scale $M_X$ implies that $\lambda_{\alpha 1}=\lambda_0\lesssim 0.8$ and $\kappa_{i 2}=\kappa_0\lesssim 0.79$.

Setting $\mu_{H_{\alpha}}=\mu_H$, $\mu_{D_i}=\mu_D$ and $M_{N_1}\simeq M_{N_2}\simeq M_{A_1}\simeq M_{A_2}\simeq M_{X}=750\,\mbox{GeV}$
one can obtain simple analytical expressions for the partial decay widths of $N_1$, $A_1$, $N_2$ and $A_2$ into a pair of photons
\begin{eqnarray}
\Gamma (N_1 \to \gamma\gamma ) =  \ds\frac{\alpha^2 M_X^3}{256 \pi^3}\Biggl|\ds\frac{\lambda_0}{\mu_H} A(x_H)
+ \frac{\kappa_0}{2 \mu_D} A(x_D) \Biggr|^2\,,\label{394}\\
\Gamma (A_1 \to \gamma\gamma )  = \ds \frac{\alpha^2 M_X^3}{256 \pi^3}\Biggl| \ds\frac{\lambda_0}{\mu_H} B(x_H)
+ \frac{\kappa_0}{2 \mu_D} B(x_D) \Biggr|^2\,,\label{395}\\
\Gamma (N_2 \to \gamma\gamma ) =  \ds \frac{\alpha^2 M_X^3}{256 \pi^3} \Biggl| \ds\frac{\lambda_0}{\mu_H} A(x_H)
- \frac{\kappa_0}{2 \mu_D} A(x_D) \Biggr|^2\,,\label{396}\\
\Gamma (A_2 \to \gamma\gamma ) =  \ds\frac{\alpha^2 M_X^3}{256 \pi^3}\Biggl| \ds\frac{\lambda_0}{\mu_H} B(x_H)
- \frac{\kappa_0}{2 \mu_D} B(x_D) \Biggr|^2\,,\label{397}
\end{eqnarray}
where $x_{D}=4\mu_{D}^2/M_X^2$ and $x_{H}=4\mu_{H}^2/M_X^2$. Assuming, that $\kappa_0/\mu_D$ and
$\lambda_0/\mu_D$ have the same sign, Eqs.~(\ref{394}) and (\ref{395}) are very similar to the ones which was used
before for the calculation of the corresponding partial decay widths in one scalar/pseudoscalar case. Because the
expressions for other partial decay widths are also very similar the branching ratios shown in Figs.~3a and 3b are
almost the same as in Figs.~1a and 1b. At the same time in the case of $N_2$ and $A_2$ destructive interference
between the contributions of exotic quarks and inert Higgsinos occurs. This leads to the suppression of the diphoton
partial decay width. As a consequence when exotic quarks are lighter than $1\,\mbox{TeV}$ the branching ratios of
the decays $N_2\to \gamma\gamma$ and $A_2\to \gamma\gamma$ are the lowest ones (see Figs.~3c and 3d).

\begin{figure}
\hspace*{0cm}{$\mbox{BR}(A_{1} \to gg, WW, ZZ, \gamma\gamma, \gamma Z)$}
\hspace*{3cm}{$\mbox{BR}(N_{1} \to gg, WW, ZZ, \gamma\gamma, \gamma Z)$}\\[-4mm]
\includegraphics[width=75mm,height=55mm]{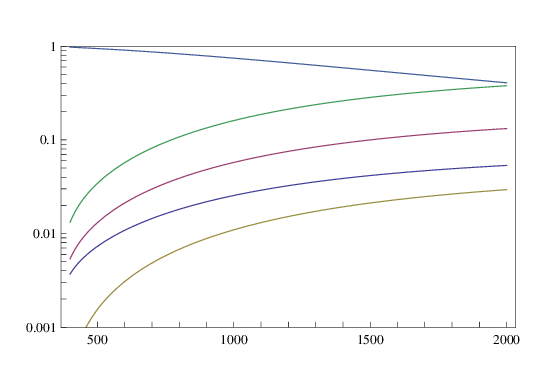}\qquad
\includegraphics[width=75mm,height=55mm]{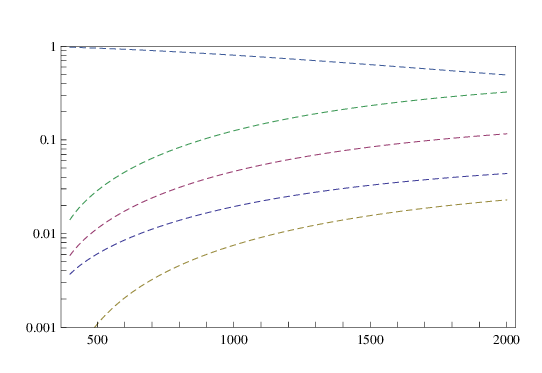}\\[-5mm]
\hspace*{3.5cm}{$\mu_D$}\hspace*{7.8cm}{$\mu_D$}\\[1mm]
\hspace*{3.5cm}{\bf (a)}\hspace*{7.5cm}{\bf (b) }\\[6mm]
\hspace*{0cm}{$\mbox{BR}(A_{2} \to gg, WW, ZZ, \gamma Z, \gamma \gamma)$}
\hspace*{3cm}{$\mbox{BR}(N_{2} \to gg, WW, ZZ, \gamma Z, \gamma \gamma)$}\\[-5mm]
\includegraphics[width=75mm,height=55mm]{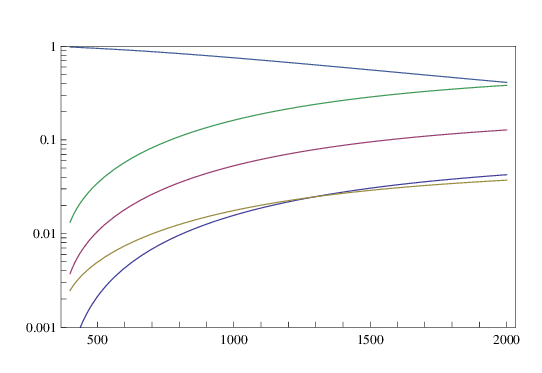}\qquad
\includegraphics[width=75mm,height=55mm]{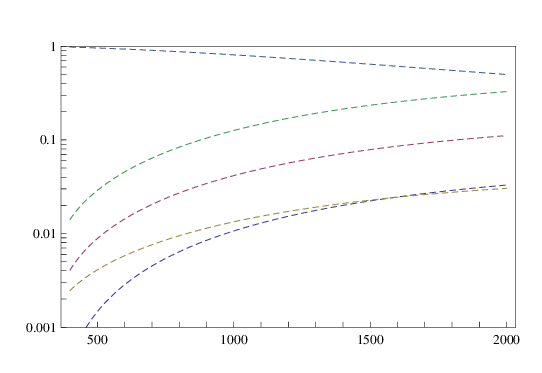}\\[-5mm]
\hspace*{3.5cm}{$\mu_D$}\hspace*{7.8cm}{$\mu_D$}\\[1mm]
\hspace*{3.5cm}{\bf (c)}\hspace*{7.5cm}{\bf (d)}
\caption{Predictions for the branching ratios in the maximal mixing scenario
for two maximally mixed
pseudoscalars (left panels) or two maximally mixed scalars (right panels) case.
In all cases the masses of exotic quarks are set to be equal, i.e. $\mu_{D_i}=\mu_D$, while $\mu_{H_{\alpha}}=400\,\mbox{GeV}$,
$\lambda_{\alpha 1}=0.8$, $\kappa_{i2}=0.79$ and $\kappa_{i1}=\lambda_{\alpha 2}=0$.
In {\it (a)} the branching ratios of the decays of $A_{1}$ into $\gamma Z$ (lowest solid line),
$\gamma\gamma$ (second lowest solid line), $ZZ$ (third lowest solid line), $WW$ (second highest solid line)
and $gg$ (highest solid line) as a function of exotic quark masses for $M_{A_{1}}\simeq 750\,\mbox{GeV}$.
In {\it (b)} the branching ratios of the decays of $N_{1}$ into $\gamma Z$ (lowest dashed line),
$\gamma\gamma$ (second lowest dashed line), $ZZ$ (third lowest dashed line), $WW$ (second highest dashed line)
and $gg$ (highest dashed line) as a function of exotic quark masses for $M_{N_{1}}\simeq 750\,\mbox{GeV}$.
In {\it (c)} the branching ratios of the decays of $A_{2}$ into $\gamma\gamma$ (lowest solid line),
$\gamma Z$ (second lowest solid line), $ZZ$ (third lowest solid line), $WW$ (second highest solid line)
and $gg$ (highest solid line) as a function of exotic quark masses for $M_{A_{1}}\simeq 750\,\mbox{GeV}$.
In {\it (d)} the branching ratios of the decays of $N_{2}$ into $\gamma \gamma$ (lowest dashed line),
$\gamma Z$ (second lowest dashed line), $ZZ$ (third lowest dashed line), $WW$ (second highest dashed line)
and $gg$ (highest dashed line) as a function of exotic quark masses for $M_{N_{1}}\simeq 750\,\mbox{GeV}$.
}
\label{fig3}
\end{figure}

As before from Fig.~3 it follows that all exotic states $N_1$, $A_1$, $N_2$ and $A_2$ decay mainly into a pair of gluons.
The corresponding branching ratio decreases with increasing $\mu_D$ because $c_{3\alpha}$ and $\tilde{c}_{3\alpha}$
diminish. The branching ratios of the decay of these states into $WW$ and $ZZ$ are the second largest and third largest
ones. These branching ratios are substantially larger than the ones associated with the decays of exotic states into
$\gamma\gamma$ and $\gamma Z$. In the case of $N_2$ and $A_2$ the branching ratios of the decay of these states
into $WW$ can be an order of magnitude larger than the branching ratios of $N_2\to \gamma\gamma$ and $A_2\to\gamma\gamma$.
Nevertheless the observation of the decays of $N_{\alpha}$ and $A_{\alpha}$ into pairs of $WW$ and $ZZ$ tend to be
more problematic since $W$ and $Z$ decay mostly into quarks. All branching ratios of the exotic scalar and pseudoscalar
decays except the largest one grow with increasing $\mu_D$. As a result for $\mu_D\simeq 2\,\mbox{TeV}$ the branching ratios
of $A_{\alpha}(N_{\alpha})\to gg$ and $A_{\alpha}(N_{\alpha})\to WW$ become sufficiently close.

The dependence of the partial decay widths and the corresponding cross sections at the $13\,\mbox{TeV}$ LHC associated with
the decays of the exotic pseudoscalar and scalar states into a pair of photons on the exotic quark masses is shown in Fig.~4.
The results of our calculations for $N_1$ and  $A_1$ are very similar to the ones obtained in the one scalar/pseudoscalar
case (see Fig. 2e and 2f). The partial decay widths and the cross sections $\sigma(pp \to A_{1}(N_1) \to \gamma\gamma)$
are just a bit smaller since the Yukawa couplings of $A_1$ and $N_1$ to the exotic quarks and inert Higgsino states are
slightly smaller. They decrease with increasing the masses of exotic quarks $\mu_D$ as before. On the contrary, the partial decay
widths of $N_2\to\gamma\gamma$ and $A_2\to \gamma\gamma$ increase with increasing the exotic quark masses for
fixed values of inert Higgsino masses because of the destructive interference mentioned above. They attain their maximal
values for $\mu_D\gg 1\,\mbox{TeV}$ when the contribution of the exotic quarks to the partial decay widths become
vanishingly small. The cross sections $\sigma(pp \to A_{2}(N_2) \to \gamma\gamma)$ also increase with increasing
exotic quark masses when $\mu_D\lesssim 700\,\mbox{GeV}$. However if exotic quarks are considerably heavier
than $1\,\mbox{TeV}$ then these cross sections become smaller for larger $\mu_D$ since the branching ratios of
$A_{2}(N_{2})\to gg$ diminish.

The sums of the cross sections $\sigma(pp \to N_1 \to \gamma\gamma)+\sigma(pp \to N_2 \to \gamma\gamma)$ and
$\sigma(pp \to A_1 \to \gamma\gamma)+$ $\sigma(pp \to A_2 \to \gamma\gamma)$ that correspond to the case when
all exotic scalar and pseudoscalar states have masses around $750\,\mbox{GeV}$ decreases with increasing $\mu_D$
(see Figs.~4c and 4d). At large values of the exotic quark masses these cross sections are bigger than the ones in
the one scalar/pseudoscalar case shown in Fig. 1e and 1f. This is because the requirement of the validity of perturbation
theory up to the scale $M_X$ allows for larger values of $\lambda_{\alpha 1}$ in the maximal mixing scenario as compared
with the one scalar/pseudoscalar case. From Figs.~4c and 4d one can see that the sum of all cross section that includes
contributions of all scalar and pseudoscalar states with masses around $750\,\mbox{GeV}$ changes from $4.5\,\mbox{fb}$
to $3\,\mbox{fb}$ when the exotic quark masses vary from $400\,\mbox{GeV}$ to $1\,\mbox{TeV}$. The presence of
such nearly degenerate states in the particle spectrum may also provide an explanation why the value of the best-fit width
of the resonance obtained by ATLAS collaboration is so large.

\begin{figure}
\hspace*{0cm}{$\frac{\Gamma(A_{1,2} \to \gamma\gamma)}{M_X} \times 10^6$}
\hspace*{5.6cm}{$\frac{\Gamma(N_{1,2} \to \gamma\gamma)}{M_X} \times 10^6$}\\[-4mm]
\includegraphics[width=75mm,height=55mm]{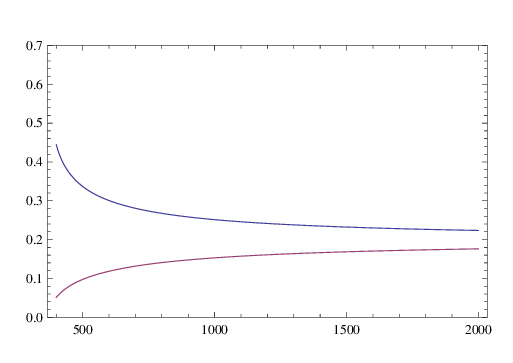}\qquad
\includegraphics[width=75mm,height=55mm]{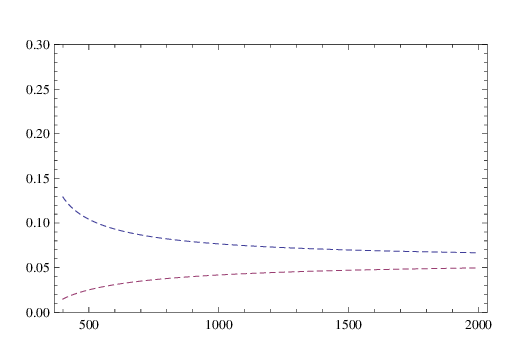}\\[-5mm]
\hspace*{3.5cm}{$\mu_D$}\hspace*{7.8cm}{$\mu_D$}\\[1mm]
\hspace*{3.5cm}{\bf (a)}\hspace*{7.5cm}{\bf (b) }\\[3mm]
\hspace*{0cm}{$\sigma(pp\to A_{1,2} \to \gamma\gamma)[\mbox{fb}]$}
\hspace*{4.7cm}{$\sigma(pp\to N_{1,2} \to \gamma\gamma)[\mbox{fb}]$}\\[-5mm]
\includegraphics[width=75mm,height=55mm]{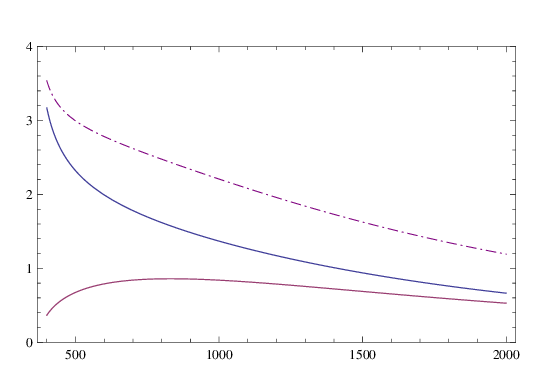}\qquad
\includegraphics[width=75mm,height=55mm]{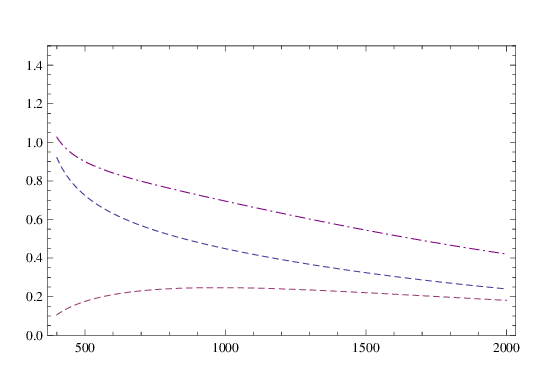}\\[-5mm]
\hspace*{3.5cm}{$\mu_D$}\hspace*{7.8cm}{$\mu_D$}\\[1mm]
\hspace*{3.5cm}{\bf (c)}\hspace*{7.5cm}{\bf (d)}
\caption{
Predictions for the maximal mixing scenario
for two maximally mixed
pseudoscalars (left panels) or two maximally mixed scalars (right panels) case.
In all cases $\mu_{H_{\alpha}}=400\,\mbox{GeV}$, $\lambda_{\alpha 1}=0.8$,
$\kappa_{i 2}=0.79$, $\lambda_{\alpha 2}=\kappa_{i 1}=0$ and
the masses of exotic quarks are set to be equal, i.e. $\mu_{D_i}=\mu_D$.
In {\it (a)} the ratios ${\Gamma(A_{1}\to \gamma\gamma)}/{M_X}$ (upper solid line)
and ${\Gamma(A_{2}\to \gamma\gamma)}/{M_X}$ (lower solid line)
as a function of exotic quark masses in the maximal mixing scenario for $M_{A_{\alpha}}\simeq 750\,\mbox{GeV}$.
In {\it (b)} the ratios ${\Gamma(N_{1}\to \gamma\gamma)}/{M_X}$ (upper dashed line)
and ${\Gamma(N_{2}\to \gamma\gamma)}/{M_X}$ (lower dashed line)
as a function of exotic quark masses in the maximal mixing scenario for $M_{N_{\alpha}}\simeq 750\,\mbox{GeV}$.
In {\it (c)} the cross sections in fb $\sigma(pp\to A_{1} \to \gamma\gamma)$ (upper solid line) and $\sigma(pp\to A_{2} \to \gamma\gamma)$ (lower solid line)
as a function of exotic quark masses for $M_{A_{1,2}}\simeq 750\,\mbox{GeV}$. The dashed--dotted line correspond to the sum of these cross sections.
In {\it (d)} the cross sections in fb $\sigma(pp\to N_{1} \to \gamma\gamma)$ (upper dashed line) and $\sigma(pp\to N_{2} \to \gamma\gamma)$ (lower dashed line)
as a function of exotic quark masses for $M_{N_{1,2}}\simeq 750\,\mbox{GeV}$. The dashed--dotted line correspond to the sum of these cross sections.
}
\label{fig4}
\end{figure}

\section{Conclusions}
\label{conclusions}

In this paper we have proposed a variant of the E$_6$SSM in which
the third singlet $S_3$ breaks the gauged $U(1)_N$ above the TeV scale,
which predicts a $Z'_N$, vector-like colour triplet and charge $\mp 1/3$ quarks
$D,\bar{D}$, and two families of inert Higgsinos, all of which should be observed
at LHC Run 2, plus
the two lighter singlets $\hat{S}_{1,2}$ with masses around 750 GeV which are
candidates for the recently observed diphoton excess.
We have calculated the branching ratios and cross-sections for the two scalars
$N_{1,2}$ and two pseudoscalars $A_{1,2}$
associated with  $\hat{S}_{1,2}$, including possible degeneracies and maximal mixing,
subject to the constraint that their couplings remain perturbative
up to the unification scale.

Our results show that this variant of the E$_6$SSM with
two nearly degenerate pseudoscalars $A_{1,2}$
with masses around $750\,\mbox{GeV}$,
may give rise to cross sections of $pp\to \gamma\gamma$ that can be as large
as about $3\,\mbox{fb}$ providing that the
inert Higgsino states have masses around $400\,\mbox{GeV}$,
while the three generations of $D,\bar{D}$ are lighter than about $1\,\mbox{TeV}$.
If the two nearly denegerate scalars $N_{1,2}$ also have
masses around $750\,\mbox{GeV}$, then these cross-sections may be further boosted
by about $1\,\mbox{fb}$, assuming that they are at present unresolvable.
The existence of nearly degenerate spinless
singlets provides an
explaination for why the best-fit width of the $750\,\mbox{GeV}$ resonance
obtained by the ATLAS collaboration is apparently so large, i.e. about $45\,\mbox{GeV}$.
However further data from Run 2 should begin to resolve the two separate
pseudoscalar states $A_{1,2}$ (plus perhaps the two scalar states $N_{1,2}$).

Finally we emphasise that
the three families of light vector-like D-quarks around 1 TeV and two families of inert
Higgsinos around 400 GeV, although
not currently ruled out because of their non--standard decay patterns,
should be observable in dedicated searches at Run 2 of the LHC.
The $Z'_N$ gauge boson also remains a prediction of the E$_6$SSM.
In addition, the proposed variant E$_6$SSM
also predicts further decay modes of the 750 GeV resonance into
$WW$, $ZZ$ and $\gamma Z$ that might be possible to observe in the Run 2 at the LHC.

\vspace{0.5cm}
\section*{Acknowledgements}
RN is grateful to P.~Athron, P.~Jackson, J.~Li, M.~M\"{u}hlleitner, P.~Sharma, R.~Young and  J.~Zanotti
for helpful discussions. RN also thanks E.~Boos, X.~Tata and A.~W.~Thomas for useful comments
and remarks. The work of R.N. was supported by the University of Adelaide and the
Australian Research Council through the ARC Center of Excellence in
Particle Physics at the Terascale.
SFK acknowledges partial support from the STFC Consolidated ST/J000396/1 grant and
the European Union FP7 ITN-INVISIBLES (Marie Curie Actions, PITN-
GA-2011-289442).


\begin{thebibliography}{99}
\bibitem{ATLAS}
ATLAS Collaboration, ``Search for resonances decaying to photon pairs in 3.2 fb$^{?1}$ of pp collisions at $\sqrt{s}$ = 13 TeV with the ATLAS detector'', ATLAS-CONF-2015- 081.

\bibitem{CMS}
CMS Collaboration, ``Search for new physics in high mass diphoton events in proton-proton collisions at $\sqrt{s}$  = 13 TeV'', CMS-PAS-EXO-15-004.



 \bibitem{Harigaya:2015ezk}
   K.~Harigaya and Y.~Nomura,
   arXiv:1512.04850 [hep-ph].


\bibitem{Mambrini:2015wyu}
  Y.~Mambrini, G.~Arcadi and A.~Djouadi,
  arXiv:1512.04913 [hep-ph].

  \bibitem{Backovic:2015fnp}
  M.~Backovic, A.~Mariotti and D.~Redigolo,
  arXiv:1512.04917 [hep-ph].


\bibitem{Angelescu:2015uiz}
  A.~Angelescu, A.~Djouadi and G.~Moreau,
  arXiv:1512.04921 [hep-ph].


\bibitem{Nakai:2015ptz}
  Y.~Nakai, R.~Sato and K.~Tobioka,
  arXiv:1512.04924 [hep-ph].


\bibitem{Knapen:2015dap}
  S.~Knapen, T.~Melia, M.~Papucci and K.~Zurek,
  arXiv:1512.04928 [hep-ph].

   \bibitem{Pilaftsis:2015ycr}
     A.~Pilaftsis,
     arXiv:1512.04931 [hep-ph].



\bibitem{Franceschini:2015kwy}
 R.~Franceschini {\it et al.},
  arXiv:1512.04933 [hep-ph].

\bibitem{DiChiara:2015vdm}
  S.~Di Chiara, L.~Marzola and M.~Raidal,
  arXiv:1512.04939 [hep-ph].

  \bibitem{Ellis:2015oso}
  J.~Ellis, S.~A.~R.~Ellis, J.~Quevillon, V.~Sanz and T.~You,
  arXiv:1512.05327 [hep-ph].

\bibitem{Bellazzini:2015nxw}
  B.~Bellazzini, R.~Franceschini, F.~Sala and J.~Serra,
  arXiv:1512.05330 [hep-ph].

  \bibitem{Gupta:2015zzs}
  R.~S.~Gupta, S.~Jäger, Y.~Kats, G.~Perez and E.~Stamou,
  arXiv:1512.05332 [hep-ph].
  \bibitem{Higaki:2015jag}
  T.~Higaki, K.~S.~Jeong, N.~Kitajima and F.~Takahashi,
  arXiv:1512.05295 [hep-ph].
  \bibitem{McDermott:2015sck}
  S.~D.~McDermott, P.~Meade and H.~Ramani,
  arXiv:1512.05326 [hep-ph].
  \bibitem{Low:2015qep}
  M.~Low, A.~Tesi and L.~T.~Wang,
  arXiv:1512.05328 [hep-ph].
  \bibitem{Petersson:2015mkr}
  C.~Petersson and R.~Torre,
  arXiv:1512.05333 [hep-ph].

\bibitem{Molinaro:2015cwg}
  E.~Molinaro, F.~Sannino and N.~Vignaroli,
  arXiv:1512.05334 [hep-ph].


  \bibitem{Dutta:2015wqh}
  B.~Dutta, Y.~Gao, T.~Ghosh, I.~Gogoladze and T.~Li,
  arXiv:1512.05439 [hep-ph].
  \bibitem{Cao:2015pto}
  Q.~H.~Cao, Y.~Liu, K.~P.~Xie, B.~Yan and D.~M.~Zhang,
  arXiv:1512.05542 [hep-ph].
  \bibitem{Kobakhidze:2015ldh}
  A.~Kobakhidze, F.~Wang, L.~Wu, J.~M.~Yang and M.~Zhang,
  arXiv:1512.05585 [hep-ph].
  \bibitem{Cox:2015ckc}
  P.~Cox, A.~D.~Medina, T.~S.~Ray and A.~Spray,
  arXiv:1512.05618 [hep-ph].
  \bibitem{Ahmed:2015uqt}
  A.~Ahmed, B.~M.~Dillon, B.~Grzadkowski, J.~F.~Gunion and Y.~Jiang,
  arXiv:1512.05771 [hep-ph].
  \bibitem{Becirevic:2015fmu}
  D.~Becirevic, E.~Bertuzzo, O.~Sumensari and R.~Z.~Funchal,
  arXiv:1512.05623 [hep-ph].
  \bibitem{No:2015bsn}
  J.~M.~No, V.~Sanz and J.~Setford,
  arXiv:1512.05700 [hep-ph].
  \bibitem{Demidov:2015zqn}
  S.~V.~Demidov and D.~S.~Gorbunov,
  arXiv:1512.05723 [hep-ph].
  \bibitem{Chao:2015ttq}
  W.~Chao, R.~Huo and J.~H.~Yu,
  arXiv:1512.05738 [hep-ph].
  \bibitem{Fichet:2015vvy}
  S.~Fichet, G.~von Gersdorff and C.~Royon,
  arXiv:1512.05751 [hep-ph].
  \bibitem{Curtin:2015jcv}
  D.~Curtin and C.~B.~Verhaaren,
  arXiv:1512.05753 [hep-ph].
  \bibitem{Bian:2015kjt}
  L.~Bian, N.~Chen, D.~Liu and J.~Shu,
  arXiv:1512.05759 [hep-ph].
  \bibitem{Chakrabortty:2015hff}
  J.~Chakrabortty, A.~Choudhury, P.~Ghosh, S.~Mondal and T.~Srivastava,
  arXiv:1512.05767 [hep-ph].
  \bibitem{Csaki:2015vek}
  C.~Csaki, J.~Hubisz and J.~Terning,
  arXiv:1512.05776 [hep-ph].
  \bibitem{Falkowski:2015swt}
  A.~Falkowski, O.~Slone and T.~Volansky,
  arXiv:1512.05777 [hep-ph].
  \bibitem{Bai:2015nbs}
  Y.~Bai, J.~Berger and R.~Lu,
  arXiv:1512.05779 [hep-ph].
  \bibitem{Benbrik:2015fyz}
  R.~Benbrik, C.~H.~Chen and T.~Nomura,
  arXiv:1512.06028 [hep-ph].
  \bibitem{Kim:2015ron}
  J.~S.~Kim, J.~Reuter, K.~Rolbiecki and R.~R.~de Austri,
  arXiv:1512.06083 [hep-ph].

  \bibitem{Gabrielli:2015dhk}
  E.~Gabrielli, K.~Kannike, B.~Mele, M.~Raidal, C.~Spethmann and H.~Veermäe,
  arXiv:1512.05961 [hep-ph].
  \bibitem{Alves:2015jgx}
  A.~Alves, A.~G.~Dias and K.~Sinha,
  arXiv:1512.06091 [hep-ph].
  \bibitem{Megias:2015ory}
  E.~Megias, O.~Pujolas and M.~Quiros,
  arXiv:1512.06106 [hep-ph].
  \bibitem{Carpenter:2015ucu}
  L.~M.~Carpenter, R.~Colburn and J.~Goodman,
  arXiv:1512.06107 [hep-ph].
  \bibitem{Bernon:2015abk}
  J.~Bernon and C.~Smith,
  arXiv:1512.06113 [hep-ph].
  \bibitem{Chao:2015nsm}
  W.~Chao,
  arXiv:1512.06297 [hep-ph].
  \bibitem{Han:2015cty}
  C.~Han, H.~M.~Lee, M.~Park and V.~Sanz,
  arXiv:1512.06376 [hep-ph].
  \bibitem{Chang:2015bzc}
  S.~Chang,
  arXiv:1512.06426 [hep-ph].
  \bibitem{Dhuria:2015ufo}
  M.~Dhuria and G.~Goswami,
  arXiv:1512.06782 [hep-ph].
  \bibitem{Han:2015dlp}
  H.~Han, S.~Wang and S.~Zheng,
  arXiv:1512.06562 [hep-ph].
  \bibitem{Luo:2015yio}
  M.~x.~Luo, K.~Wang, T.~Xu, L.~Zhang and G.~Zhu,
  arXiv:1512.06670 [hep-ph].
  \bibitem{Chang:2015sdy}
  J.~Chang, K.~Cheung and C.~T.~Lu,
  arXiv:1512.06671 [hep-ph].
  \bibitem{Bardhan:2015hcr}
  D.~Bardhan, D.~Bhatia, A.~Chakraborty, U.~Maitra, S.~Raychaudhuri and T.~Samui,
  arXiv:1512.06674 [hep-ph].
  \bibitem{Feng:2015wil}
  T.~F.~Feng, X.~Q.~Li, H.~B.~Zhang and S.~M.~Zhao,
  arXiv:1512.06696 [hep-ph].
  \bibitem{Liao:2015tow}
  W.~Liao and H.~q.~Zheng,
  arXiv:1512.06741 [hep-ph].
  \bibitem{Cho:2015nxy}
  W.~S.~Cho, D.~Kim, K.~Kong, S.~H.~Lim, K.~T.~Matchev, J.~C.~Park and M.~Park,
  arXiv:1512.06824 [hep-ph].
  \bibitem{Barducci:2015gtd}
  D.~Barducci, A.~Goudelis, S.~Kulkarni and D.~Sengupta,
  arXiv:1512.06842 [hep-ph].
  \bibitem{Ding:2015rxx}
  R.~Ding, L.~Huang, T.~Li and B.~Zhu,
  arXiv:1512.06560 [hep-ph].
  \bibitem{Han:2015qqj}
  X.~F.~Han and L.~Wang,
  arXiv:1512.06587 [hep-ph].
  \bibitem{Antipin:2015kgh}
  O.~Antipin, M.~Mojaza and F.~Sannino,
  arXiv:1512.06708 [hep-ph].
  \bibitem{Wang:2015kuj}
  F.~Wang, L.~Wu, J.~M.~Yang and M.~Zhang,
  arXiv:1512.06715 [hep-ph].
  \bibitem{Cao:2015twy}
  J.~Cao, C.~Han, L.~Shang, W.~Su, J.~M.~Yang and Y.~Zhang,
  arXiv:1512.06728 [hep-ph].
  \bibitem{Huang:2015evq}
  F.~P.~Huang, C.~S.~Li, Z.~L.~Liu and Y.~Wang,
  arXiv:1512.06732 [hep-ph].
  \bibitem{Heckman:2015kqk}
  J.~J.~Heckman,
  arXiv:1512.06773 [hep-ph].
  \bibitem{Bi:2015uqd}
  X.~J.~Bi, Q.~F.~Xiang, P.~F.~Yin and Z.~H.~Yu,
  arXiv:1512.06787 [hep-ph].

  \bibitem{Kim:2015ksf}
  J.~S.~Kim, K.~Rolbiecki and R.~R.~de Austri,
  arXiv:1512.06797 [hep-ph].
  \bibitem{Berthier:2015vbb}
  L.~Berthier, J.~M.~Cline, W.~Shepherd and M.~Trott,
  arXiv:1512.06799 [hep-ph].
  \bibitem{Cline:2015msi}
  J.~M.~Cline and Z.~Liu,
  arXiv:1512.06827 [hep-ph].
  \bibitem{Bauer:2015boy}
  M.~Bauer and M.~Neubert,
  arXiv:1512.06828 [hep-ph].
  \bibitem{Chala:2015cev}
  M.~Chala, M.~Duerr, F.~Kahlhoefer and K.~Schmidt-Hoberg,
  arXiv:1512.06833 [hep-ph].
  \bibitem{Boucenna:2015pav}
  S.~M.~Boucenna, S.~Morisi and A.~Vicente,
  arXiv:1512.06878 [hep-ph].
  \bibitem{Dev:2015isx}
  P.~S.~B.~Dev and D.~Teresi,
  arXiv:1512.07243 [hep-ph].
  \bibitem{deBlas:2015hlv}
  J.~de Blas, J.~Santiago and R.~Vega-Morales,
  arXiv:1512.07229 [hep-ph].
  \bibitem{Murphy:2015kag}
  C.~W.~Murphy,
  arXiv:1512.06976 [hep-ph].
  \bibitem{Hernandez:2015ywg}
  A.~E.~C.~Hernández and I.~Nisandzic,
  arXiv:1512.07165 [hep-ph].
  \bibitem{Dey:2015bur}
  U.~K.~Dey, S.~Mohanty and G.~Tomar,
  arXiv:1512.07212 [hep-ph].
  \bibitem{Pelaggi:2015knk}
  G.~M.~Pelaggi, A.~Strumia and E.~Vigiani,
  arXiv:1512.07225 [hep-ph].
  \bibitem{Belyaev:2015hgo}
  A.~Belyaev, G.~Cacciapaglia, H.~Cai, T.~Flacke, A.~Parolini and H.~Serôdio,
  arXiv:1512.07242 [hep-ph].
  \bibitem{Huang:2015rkj}
  W.~C.~Huang, Y.~L.~S.~Tsai and T.~C.~Yuan,
  arXiv:1512.07268 [hep-ph].
  \bibitem{Cao:2015xjz}
  Q.~H.~Cao, S.~L.~Chen and P.~H.~Gu,
  arXiv:1512.07541 [hep-ph].
  \bibitem{Gu:2015lxj}
  J.~Gu and Z.~Liu,
  arXiv:1512.07624 [hep-ph].
  \bibitem{Patel:2015ulo}
  K.~M.~Patel and P.~Sharma,
  arXiv:1512.07468 [hep-ph].
  \bibitem{Badziak:2015zez}
  M.~Badziak,
  arXiv:1512.07497 [hep-ph].
  \bibitem{Chakraborty:2015gyj}
  S.~Chakraborty, A.~Chakraborty and S.~Raychaudhuri,
  arXiv:1512.07527 [hep-ph].
  \bibitem{Altmannshofer:2015xfo}
  W.~Altmannshofer, J.~Galloway, S.~Gori, A.~L.~Kagan, A.~Martin and J.~Zupan,
  arXiv:1512.07616 [hep-ph].
  \bibitem{Cvetic:2015vit}
  M.~Cvetic, J.~Halverson and P.~Langacker,
  arXiv:1512.07622 [hep-ph].
  \bibitem{Allanach:2015ixl}
  B.~C.~Allanach, P.~S.~B.~Dev, S.~A.~Renner and K.~Sakurai,
  arXiv:1512.07645 [hep-ph].
  \bibitem{Davoudiasl:2015cuo}
  H.~Davoudiasl and C.~Zhang,
  arXiv:1512.07672 [hep-ph].
  \bibitem{Das:2015enc}
  K.~Das and S.~K.~Rai,
  arXiv:1512.07789 [hep-ph].
  \bibitem{Cheung:2015cug}
  K.~Cheung, P.~Ko, J.~S.~Lee, J.~Park and P.~Y.~Tseng,
  arXiv:1512.07853 [hep-ph].

  \bibitem{Craig:2015lra}
  N.~Craig, P.~Draper, C.~Kilic and S.~Thomas,
  arXiv:1512.07733 [hep-ph].
  \bibitem{Liu:2015yec}
  J.~Liu, X.~P.~Wang and W.~Xue,
  arXiv:1512.07885 [hep-ph].
  \bibitem{Zhang:2015uuo}
  J.~Zhang and S.~Zhou,
  arXiv:1512.07889 [hep-ph].
  \bibitem{Casas:2015blx}
  J.~A.~Casas, J.~R.~Espinosa and J.~M.~Moreno,
  arXiv:1512.07895 [hep-ph].
  \bibitem{Hall:2015xds}
  L.~J.~Hall, K.~Harigaya and Y.~Nomura,
  arXiv:1512.07904 [hep-ph].
  \bibitem{Park:2015ysf}
  J.~C.~Park and S.~C.~Park,
  arXiv:1512.08117 [hep-ph].
  \bibitem{Salvio:2015jgu}
  A.~Salvio and A.~Mazumdar,
  arXiv:1512.08184 [hep-ph].
  \bibitem{Li:2015jwd}
  G.~Li, Y.~n.~Mao, Y.~L.~Tang, C.~Zhang, Y.~Zhou and S.~h.~Zhu,
  arXiv:1512.08255 [hep-ph].
  \bibitem{Son:2015vfl}
  M.~Son and A.~Urbano,
  arXiv:1512.08307 [hep-ph].
  \bibitem{An:2015cgp}
  H.~An, C.~Cheung and Y.~Zhang,
  arXiv:1512.08378 [hep-ph].
  \bibitem{Wang:2015omi}
  F.~Wang, W.~Wang, L.~Wu, J.~M.~Yang and M.~Zhang,
  arXiv:1512.08434 [hep-ph].
  \bibitem{Cao:2015scs}
  Q.~H.~Cao, Y.~Liu, K.~P.~Xie, B.~Yan and D.~M.~Zhang,
  arXiv:1512.08441 [hep-ph].
  \bibitem{Gao:2015igz}
  J.~Gao, H.~Zhang and H.~X.~Zhu,
  arXiv:1512.08478 [hep-ph].
  \bibitem{Goertz:2015nkp}
  F.~Goertz, J.~F.~Kamenik, A.~Katz and M.~Nardecchia,
  arXiv:1512.08500 [hep-ph].
  \bibitem{Dev:2015vjd}
  P.~S.~B.~Dev, R.~N.~Mohapatra and Y.~Zhang,
  arXiv:1512.08507 [hep-ph].
  \bibitem{Cao:2015apa}
  J.~Cao, F.~Wang and Y.~Zhang,
  arXiv:1512.08392 [hep-ph].
  \bibitem{Cai:2015hzc}
  C.~Cai, Z.~H.~Yu and H.~H.~Zhang,
  arXiv:1512.08440 [hep-ph].
  \bibitem{Kim:2015xyn}
  J.~E.~Kim,
  arXiv:1512.08467 [hep-ph].
  \bibitem{Chao:2015nac}
  W.~Chao,
  arXiv:1512.08484 [hep-ph].
  \bibitem{Anchordoqui:2015jxc}
  L.~A.~Anchordoqui, I.~Antoniadis, H.~Goldberg, X.~Huang, D.~Lust and T.~R.~Taylor,
  arXiv:1512.08502 [hep-ph].
  \bibitem{Bizot:2015qqo}
  N.~Bizot, S.~Davidson, M.~Frigerio and J.-L.~Kneur,
  arXiv:1512.08508 [hep-ph].
  \bibitem{Ibanez:2015uok}
  L.~E.~Ibanez and V.~Martin-Lozano,
  arXiv:1512.08777 [hep-ph].
  \bibitem{Huang:2015svl}
  X.~J.~Huang, W.~H.~Zhang and Y.~F.~Zhou,
  arXiv:1512.08992 [hep-ph].
  \bibitem{Chiang:2015tqz}
  C.~W.~Chiang, M.~Ibe and T.~T.~Yanagida,
  arXiv:1512.08895 [hep-ph].
  \bibitem{Kang:2015roj}
  S.~K.~Kang and J.~Song,
  arXiv:1512.08963 [hep-ph].

  \bibitem{Kanemura:2015bli}
  S.~Kanemura, K.~Nishiwaki, H.~Okada, Y.~Orikasa, S.~C.~Park and R.~Watanabe,
  arXiv:1512.09048 [hep-ph].
  \bibitem{Low:2015qho}
  I.~Low and J.~Lykken,
  arXiv:1512.09089 [hep-ph].
  \bibitem{Hernandez:2015hrt}
  A.~E.~C.~Hernández,
  arXiv:1512.09092 [hep-ph].
  \bibitem{Kaneta:2015qpf}
  K.~Kaneta, S.~Kang and H.~S.~Lee,
  arXiv:1512.09129 [hep-ph].
  \bibitem{Dasgupta:2015pbr}
  A.~Dasgupta, M.~Mitra and D.~Borah,
  arXiv:1512.09202 [hep-ph].


\bibitem{Berlin:2016hqw}
  A.~Berlin,
  arXiv:1601.01381 [hep-ph].

\bibitem{Zhang:2016pip}
  H.~Zhang,
  arXiv:1601.01355 [hep-ph].

\bibitem{Deppisch:2016scs}
  F.~F.~Deppisch, C.~Hati, S.~Patra, P.~Pritimita and U.~Sarkar,
  arXiv:1601.00952 [hep-ph].

\bibitem{Dutta:2016jqn}
  B.~Dutta, Y.~Gao, T.~Ghosh, I.~Gogoladze, T.~Li, Q.~Shafi and J.~W.~Walker,
  arXiv:1601.00866 [hep-ph].

\bibitem{Modak:2016ung}
  T.~Modak, S.~Sadhukhan and R.~Srivastava,
  arXiv:1601.00836 [hep-ph].

\bibitem{Hernandez:2016rbi}
  A.~E.~C.~Hern‡ndez, I.~d.~M.~Varzielas and E.~Schumacher,
  arXiv:1601.00661 [hep-ph].

\bibitem{Csaki:2016raa}
  C.~Csaki, J.~Hubisz, S.~Lombardo and J.~Terning,
  arXiv:1601.00638 [hep-ph].

\bibitem{Chao:2016mtn}
  W.~Chao,
  arXiv:1601.00633 [hep-ph].

\bibitem{Danielsson:2016nyy}
  U.~Danielsson, R.~Enberg, G.~Ingelman and T.~Mandal,
  arXiv:1601.00624 [hep-ph].


\bibitem{Ghorbani:2016jdq}
  K.~Ghorbani and H.~Ghorbani,
  arXiv:1601.00602 [hep-ph].


\bibitem{Han:2016bus}
  X.~F.~Han, L.~Wang, L.~Wu, J.~M.~Yang and M.~Zhang,
  arXiv:1601.00534 [hep-ph].

\bibitem{Ko:2016lai}
  P.~Ko, Y.~Omura and C.~Yu,
  arXiv:1601.00586 [hep-ph].

\bibitem{Nomura:2016fzs}
  T.~Nomura and H.~Okada,
  arXiv:1601.00386 [hep-ph].


\bibitem{Palti:2016kew}
  E.~Palti,
  arXiv:1601.00285 [hep-ph].

\bibitem{Potter:2016psi}
  C.~T.~Potter,
  arXiv:1601.00240 [hep-ph].



\bibitem{Bhattacharya:2016lyg}
S.~Bhattacharya, S.~Patra, N.~Sahoo and N.~Sahu,
arXiv:1601.01569 [hep-ph].


\bibitem{Borah:2016uoi}
D.~Borah, S.~Patra and S.~Sahoo,
arXiv:1601.01828 [hep-ph].

\bibitem{Ko:2016wce}
P.~Ko and T.~Nomura,
arXiv:1601.02490 [hep-ph].


\bibitem{Cao:2016udb}
J.~Cao, L.~Shang, W.~Su, Y.~Zhang and J.~Zhu,
arXiv:1601.02570 [hep-ph].

\bibitem{Fabbrichesi:2016alj}
M.~Fabbrichesi and A.~Urbano,
arXiv:1601.02447 [hep-ph].

\bibitem{Hati:2016thk}
C.~Hati,
arXiv:1601.02457 [hep-ph].

\bibitem{Yu:2016lof}
J.~H.~Yu,
arXiv:1601.02609 [hep-ph].

\bibitem{Ding:2016ldt}
R.~Ding, Z.~L.~Han, Y.~Liao and X.~D.~Ma,
arXiv:1601.02714 [hep-ph].


\bibitem{Dorsner:2016ypw}
I.~Dorsner, S.~Fajfer and N.~Kosnik,
arXiv:1601.03267 [hep-ph].


\bibitem{Djouadi:2016eyy}
A.~Djouadi, J.~Ellis, R.~Godbole and J.~Quevillon,
arXiv:1601.03696 [hep-ph].

\bibitem{Faraggi:2016xnm}
A.~E.~Faraggi and J.~Rizos,
arXiv:1601.03604 [hep-ph].

\bibitem{Ghoshal:2016jyj}
A.~Ghoshal,
arXiv:1601.04291 [hep-ph].

\bibitem{Nomura:2016seu}
T.~Nomura and H.~Okada,
arXiv:1601.04516 [hep-ph].

W.~Chao,
arXiv:1601.04678 [hep-ph].

\bibitem{Han:2016bvl}
X.~F.~Han, L.~Wang and J.~M.~Yang,
arXiv:1601.04954 [hep-ph].

\bibitem{Okada:2016rav}
H.~Okada and K.~Yagyu,
arXiv:1601.05038 [hep-ph].

\bibitem{Franzosi:2016wtl}
D.~B.~Franzosi and M.~T.~Frandsen,
arXiv:1601.05357 [hep-ph].

\bibitem{Martini:2016ahj}
A.~Martini, K.~Mawatari and D.~Sengupta,
arXiv:1601.05729 [hep-ph].

\bibitem{Cao:2016cok}
Q.~H.~Cao, Y.~Q.~Gong, X.~Wang, B.~Yan and L.~L.~Yang,
arXiv:1601.06374 [hep-ph].

\bibitem{Chiang:2016ydx}
C.~W.~Chiang and A.~L.~Kuo,
arXiv:1601.06394 [hep-ph].

\bibitem{Aydemir:2016qqj}
U.~Aydemir and T.~Mandal,
arXiv:1601.06761 [hep-ph].


\bibitem{Marzola:2015xbh}
L.~Marzola, A.~Racioppi, M.~Raidal, F.~R.~Urban and H.~Veermae,
arXiv:1512.09136 [hep-ph].

\bibitem{Hamada:2015skp}
Y.~Hamada, T.~Noumi, S.~Sun and G.~Shiu,
arXiv:1512.08984 [hep-ph].

\bibitem{Bi:2015lcf}
X.-J.~Bi, R.~Ding, Y.~Fan, L.~Huang, C.~Li, T.~Li, S.~Raza, X.-C.~Wang, B.~Zhu,
arXiv:1512.08497 [hep-ph].

\bibitem{Fichet:2016pvq}
S.~Fichet, G.~von Gersdorff and C.~Royon,
arXiv:1601.01712 [hep-ph].

\bibitem{Karozas:2016hcp}
  A.~Karozas, S.~F.~King, G.~K.~Leontaris and A.~K.~Meadowcroft,
  arXiv:1601.00640 [hep-ph].




\bibitem{King:2005jy}
  S.~F.~King, S.~Moretti and R.~Nevzorov,
  Phys.\ Rev.\  D {\bf 73}, 035009 (2006)
  [arXiv:hep-ph/0510419].
\bibitem{King:2005my}
  S.~F.~King, S.~Moretti and R.~Nevzorov,
  Phys.\ Lett.\  B {\bf 634}, 278 (2006)
  [arXiv:hep-ph/0511256].




\bibitem{King:2007uj}
  S.~F.~King, S.~Moretti and R.~Nevzorov,
  Phys.\ Lett.\ B {\bf 650} (2007) 57
  [hep-ph/0701064].


\bibitem{Athron:2009bs}
  P.~Athron, S.~F.~King, D.~J.~Miller, S.~Moretti and R.~Nevzorov,
  Phys.\ Rev.\  D {\bf 80} (2009) 035009
  [arXiv:0904.2169 [hep-ph]].
\bibitem{Athron:2009ue}
  P.~Athron, S.~F.~King, D.~J.~Miller, S.~Moretti, R.~Nevzorov and R.~Nevzorov,
  Phys.\ Lett.\  B {\bf 681}, 448 (2009)
  [arXiv:0901.1192 [hep-ph]].

\bibitem{Athron:2011wu}
  P.~Athron, S.~F.~King, D.~J.~Miller, S.~Moretti and R.~Nevzorov,
  Phys.\ Rev.\  D {\bf 84}, 055006 (2011)
  [arXiv:1102.4363 [hep-ph]].








\bibitem{Accomando:2006ga}
S.~F.~King, S.~Moretti, R.~Nevzorov,
arXiv:hep-ph/0601269;
S.~F.~King, S.~Moretti, R.~Nevzorov,
AIP Conf.\ Proc.\  {\bf 881} (2007) 138;
P.~Athron, S.~F.~King, D.~J.~Miller, S.~Moretti, R.~Nevzorov,
arXiv:0810.0617 [hep-ph];
P.~Athron, J.~P.~Hall, R.~Howl, S.~F.~King, D.~J.~Miller, S.~Moretti, R.~Nevzorov,
Nucl.\ Phys.\ Proc.\ Suppl.\  {\bf 200-202} (2010) 120.

\bibitem{Hall:2010ix}
S.~F.~King, R.~Luo, D.~J.~Miller, R.~Nevzorov,
JHEP {\bf 0812} (2008) 042
[arXiv:0806.0330 [hep-ph]];
J.~P.~Hall, S.~F.~King, R.~Nevzorov, S.~Pakvasa, M.~Sher,
Phys.\ Rev.\ D {\bf 83} (2011) 075013
[arXiv:1012.5114 [hep-ph]];
R.~Nevzorov,
Phys.\ Rev.\ D {\bf 87} (2013) 015029
[arXiv:1205.5967 [hep-ph]];
P.~Athron, S.~F.~King, D.~J.~Miller, S.~Moretti, R.~Nevzorov,
Phys.\ Rev.\ D {\bf 86} (2012) 095003
[arXiv:1206.5028 [hep-ph]];
P.~Athron, D.~Stockinger,  A.~Voigt,
Phys.\ Rev.\ D {\bf 86} (2012) 095012
[arXiv:1209.1470 [hep-ph]];
P.~Athron, M.~Binjonaid, S.~F.~King,
Phys.\ Rev.\ D {\bf 87}, 115023 (2013)
[arXiv:1302.5291 [hep-ph]];
M.~Sperling, D.~Stockinger, A.~Voigt,
JHEP {\bf 1307} (2013) 132
[arXiv:1305.1548 [hep-ph]];
R.~Nevzorov, S.~Pakvasa,
Phys.\ Lett.\ B {\bf 728} (2014) 210
[arXiv:1308.1021 [hep-ph]];
R.~Nevzorov,
Phys.\ Rev.\ D {\bf 89} (2014) 055010
[arXiv:1309.4738 [hep-ph]];
M.~Sperling, D.~Stockinger, A.~Voigt,
JHEP {\bf 1401}, 068 (2014)
[arXiv:1310.7629 [hep-ph]];
P.~Athron, M.~Muhlleitner, R.~Nevzorov, A.~G.~Williams,
JHEP {\bf 1501} (2015) 153
[arXiv:1410.6288 [hep-ph]];
P.~Athron, D.~Harries, R.~Nevzorov and A.~G.~Williams,
arXiv:1512.07040 [hep-ph].

\bibitem{Hall:2011zq}
  J.~P.~Hall and S.~F.~King,
  JHEP {\bf 1106} (2011) 006
  doi:10.1007/JHEP06(2011)006
  [arXiv:1104.2259 [hep-ph]];
  J.~P.~Hall and S.~F.~King,
  JHEP {\bf 0908} (2009) 088
  doi:10.1088/1126-6708/2009/08/088
  [arXiv:0905.2696 [hep-ph]];
  J.~P.~Hall and S.~F.~King,
  JHEP {\bf 1301} (2013) 076
  doi:10.1007/JHEP01(2013)076
  [arXiv:1209.4657 [hep-ph]].



\bibitem{Howl:2007zi}
  R.~Howl and S.~F.~King,
  JHEP {\bf 0801} (2008) 030
  doi:10.1088/1126-6708/2008/01/030
  [arXiv:0708.1451 [hep-ph]];
  R.~Howl and S.~F.~King,
  Phys.\ Lett.\ B {\bf 652} (2007) 331
  doi:10.1016/j.physletb.2007.07.035
  [arXiv:0705.0301 [hep-ph]];
  R.~Howl and S.~F.~King,
  JHEP {\bf 0805} (2008) 008
  doi:10.1088/1126-6708/2008/05/008
  [arXiv:0802.1909 [hep-ph]];
  R.~Howl and S.~F.~King,
  Phys.\ Lett.\ B {\bf 687} (2010) 355
  doi:10.1016/j.physletb.2010.03.053
  [arXiv:0908.2067 [hep-ph]].

\bibitem{Callaghan:2011jj}
  J.~C.~Callaghan, S.~F.~King, G.~K.~Leontaris and G.~G.~Ross,
  JHEP {\bf 1204} (2012) 094
  doi:10.1007/JHEP04(2012)094
  [arXiv:1109.1399 [hep-ph]];
  J.~C.~Callaghan and S.~F.~King,
  JHEP {\bf 1304} (2013) 034
  doi:10.1007/JHEP04(2013)034
  [arXiv:1210.6913 [hep-ph]].
  J.~C.~Callaghan, S.~F.~King and G.~K.~Leontaris,
  JHEP {\bf 1312} (2013) 037
  doi:10.1007/JHEP12(2013)037
  [arXiv:1307.4593 [hep-ph]].


\bibitem{Belyaev:2012jz}
  A.~Belyaev, J.~P.~Hall, S.~F.~King and P.~Svantesson,
  Phys.\ Rev.\ D {\bf 87} (2013) 3,  035019
  doi:10.1103/PhysRevD.87.035019
  [arXiv:1211.1962 [hep-ph]];
  A.~Belyaev, J.~P.~Hall, S.~F.~King and P.~Svantesson,
  Phys.\ Rev.\ D {\bf 86} (2012) 031702
  doi:10.1103/PhysRevD.86.031702
  [arXiv:1203.2495 [hep-ph]].
  A.~Belyaev, J.~P.~Hall, S.~F.~King and P.~Svantesson,
  Phys.\ Rev.\ D {\bf 86} (2012) 031702
  doi:10.1103/PhysRevD.86.031702
  [arXiv:1203.2495 [hep-ph]].


\bibitem{CMS-1}
The CMS Collaboration,
``Search for a Narrow Resonance Produced in 13 TeV pp Collisions Decaying to Electron Pair or Muon Pair Final States'',
CMS-PAS-EXO-15-005.


\bibitem{Frere:1996gb}
J.~M.~Frere, R.~B.~Nevzorov and M.~I.~Vysotsky,
Phys.\ Lett.\  B {\bf 394} (1997) 127
[arXiv:hep-ph/9608266].


\bibitem{Nevzorov:2001vj}
R.~Nevzorov, M.~Trusov,
Phys.\ Atom.\ Nucl.\  {\bf 64} (2001) 1299
[hep-ph/0110363];
R.~Nevzorov, M.~Trusov,
Phys.\ Atom.\ Nucl.\  {\bf 65} (2002) 335
[hep-ph/0301179];
R.~Nevzorov, M.~Trusov,
Phys.\ Atom.\ Nucl.\  {\bf 64} (2001) 1513
[hep-ph/0112301].


\end{thebibliography}
\end{document}